\documentclass[a4paper, 12pt]{article}  
\usepackage{pdfpages}
\usepackage{lmodern}
\usepackage{graphicx} 
\usepackage{wrapfig} 
\usepackage{textcomp} 
\usepackage[utf8]{inputenc}
\DeclareSymbolFont{letters}{OML}{ztmcm}{m}{it}
\usepackage{color}
\usepackage{bm}
\usepackage{multirow}
\usepackage{listings}
\usepackage{adjustbox}
\usepackage[boxed]{algorithm2e}
\usepackage{subcaption}
\usepackage{multirow} 
\usepackage[OMLmathrm,OMLmathbf]{isomath}
\usepackage{booktabs} 
\usepackage[hang,flushmargin]{footmisc}
\usepackage{pdflscape}
\usepackage{natbib}

\usepackage[T1]{fontenc} 
\usepackage{geometry}
\usepackage{amsmath}
\usepackage{etoolbox} 
\usepackage{amsthm}
\usepackage{amsfonts}
\usepackage{amssymb}
\usepackage[section]{placeins}
\usepackage{diagbox} 
\usepackage{hyperref}
\usepackage{graphicx} 
\usepackage{fixmath} 
\usepackage[scaled=.90]{helvet}
\usepackage{wrapfig}
\usepackage{tikz}

\makeatletter
\renewcommand\@biblabel[1]{\textbf{#1.}} 
\usepackage{enumitem}
\newcommand{\blind}{1}
\usepackage{pifont}
\setlist{nolistsep,leftmargin=*}
\renewcommand{\maketitle}{ 
	\begin{center}
		{\LARGE\@title} 
		
		\vspace{0pt} 
		
		{\large\@author} 
		\\\@date 
		
		\vspace{40pt} 
	\end{center}
}
\begin{document}
	
	\def\spacingset#1{\renewcommand{\baselinestretch}%
		{#1}\small\normalsize} \spacingset{1}

	
	\if1\blind
	{
		\title{\textbf{On the Interplay of Regional Mobility, Social Connectedness, and the Spread of COVID-19 in Germany}}
		\author{Cornelius Fritz and Göran Kauerman\hspace{.2cm}\\
			Department of Statistics, LMU Munich}
		\maketitle
	} \fi

	\if0\blind
	{
		\bigskip
		\bigskip
		\bigskip
		\begin{center}
			{\LARGE\bf Dynamic Networks}
		\end{center}
		\medskip
	} \fi
	
	\bigskip
	
	\begin{abstract}
Since the primary mode of respiratory virus transmission is person-to-person interaction, we are required to reconsider physical interaction patterns to mitigate the number of people infected with COVID-19. While research has shown that non-pharmaceutical interventions (NPI) had an evident impact on national mobility patterns, we investigate the relative regional mobility behaviour to assess the effect of human movement on the spread of COVID-19. In particular, we explore the impact of human mobility and social connectivity derived from Facebook activities on the weekly rate of new infections in Germany between March 3rd and June 22nd, 2020. Our results confirm that reduced social activity lowers the infection rate, accounting for regional and temporal patterns. The extent of social distancing, quantified by the percentage of people staying put within a federal administrative district, has an overall negative effect on the incidence of infections. Additionally, our results show spatial infection patterns based on geographic as well as social distances.

	\end{abstract}
	\vfill
	
	
	\section{Introduction}
	
	The COVID-19 virus outbreak originating in mainland China leapt over to Europe and quickly evolved to a global pandemic in March 2020. Only through strict non-pharmaceutical interventions (NPI) could most national health systems rapidly react to this new threat. In numerous scientific efforts,  physical distancing measures were discovered to be the most effective interventions \citep{Prem2020} and found to be necessary maybe until 2022 \citep{Kissler2020}.  The measures' effectiveness emanates from researchers confirming that the main form of virus transmission is person-to-person interaction \citep{Chan2020}. The virus can be spread by inhaling microscopic aerosol particles that contain COVID-19 and remain viable in the air with a half-life of about 1 hour \citep{Asadi2020} or direct contact through the exchange of virus-containing droplets with infected individuals \citep{Guan2020}. Since also, a high proportion of cases is asymptomatic \citep{Lavezzo2020a} and gets infected by cases in the presymptomatic stage \citep{Li2020b}, human mobility can explain the spread of COVID-19 to a considerable extent \citep{Kraemer2020}.

	Stemming from the consequential need to account for contact patterns when investigating the spread of COVID-19,  \citet{Oliver2020} list multiple possibilities of how one may utilise mobile phone data to do so. To enable this type of research, Facebook extended the \textsl{Data for Good} program to a broader audience of researchers and provided so-called \textsl{Disease Prevention Maps} for multiple countries \citep{Maas}. This database includes measurements on quantities like co-location, user counts, and movement ranges on a regional level derived from information of more than 26 million Facebook users.  Additionally, a measure for the social connectedness between geographic regions is supplied \citep{Bailey2018a}. In various studies, this data source was employed to demonstrate  how the impact of lockdown measures in Italy was more severe for municipalities with higher fiscal capacities \citep{Bonaccorsi2020a}, quantify social and geographic spillover effects from relaxations of shelter-in-place orders \citep{Holtz} and predict the number of infections on a granular spatio-temporal resolution using contact tracing data \citep{Lorch2020}.
	
	This article uses the same data source to analyse how regional differences in mobility patterns and friendship proximity affect the spread of COVID-19 in Germany.  While NPIs, e.g., the nationwide shutdown in Germany that started March 22nd, had an evident impact on national human mobility and ceased the exponential spread of the virus \citep{Flaxman2020}, the effect of the relative movement between regional districts was not yet fully assessed. So far, studies concerning human movements during the current pandemic are focused mainly on how the lockdown affected national human mobility \citep{Galeazzi2020} or specific regions regarding their economic status \citep{Bonaccorsi2020a}.
	To fill this gap, we derive covariates from the mobility data to quantify the overall dispersion of meeting patterns and compliance with social distancing.
	Through weekly standardisation of the covariates, we control for the dynamics therein, which are, in turn, driven by NPIs. As a result, our research enables a quantitative assessment of different mobility strategies relative to the national average.  Also, we infer positions of the federal administrative regions in a social space from the information on the relative friendship links among them using multidimensional scaling \citep{Cox2000}. Subsequently, we relate the processed data to Germany's weekly rate of local COVID-19 infections between March 3rd and June 22nd, 2020. This time frame permits the analysis of the dynamic spread starting with the WHO declaring COVID-19 a pandemic \citep{Organization2020}. 
	
	We employ a spatio-temporal regression model for the ratio of local COVID-19 infections that takes autoregressive structures, age and gender-specific effects, contagion by nearby districts in the geographic and social space, as well as latent heterogeneities between the districts into account. Our method is closely related to the surveillance model introduced by \citet{Held2005}. They extend generalised linear models to analyse surveillance data from epidemic outbreaks. This approach was expanded to handle multivariate surveillance data \citep{Paul2008}, control for seasonality and spatial heterogeneity \citep{Held2012}, and include neighbourhood information from social contact data \citep{Meyer2017}. In contrast to this type of model, our model's objective is to investigate the connection between mobility patterns, social connectivity, and the spread of COVID-19 in an interpretable manner. While forecasting infections is undoubtedly a central objective in epidemic surveillance, this is not the main focus of our work (see also \citealp{Held2017}). 
	
	
	The rest of the article will be structured as follows: We discuss the data sources, its measures on social interaction as well as mobility in Section \ref{sec:data}. In Section  \ref{sec:model} we detail our proposed modelling approach. We propose an imputation model for missing onset dates and use a semiparametric spatio-temporal model to analyse the ratio of local COVID-19 infections with a specific disease onset date. The results of the analysis are presented in Section \ref{sec:results}. Section \ref{sec:conclusion} concludes the article. 
	
	\section{Data description}
	\label{sec:data}
	\subsection{Data on infections}
	\label{sec:data_inf}
	
	\begin{figure}[t!]
		\centering
		\includegraphics[width=0.9\linewidth, page =1]{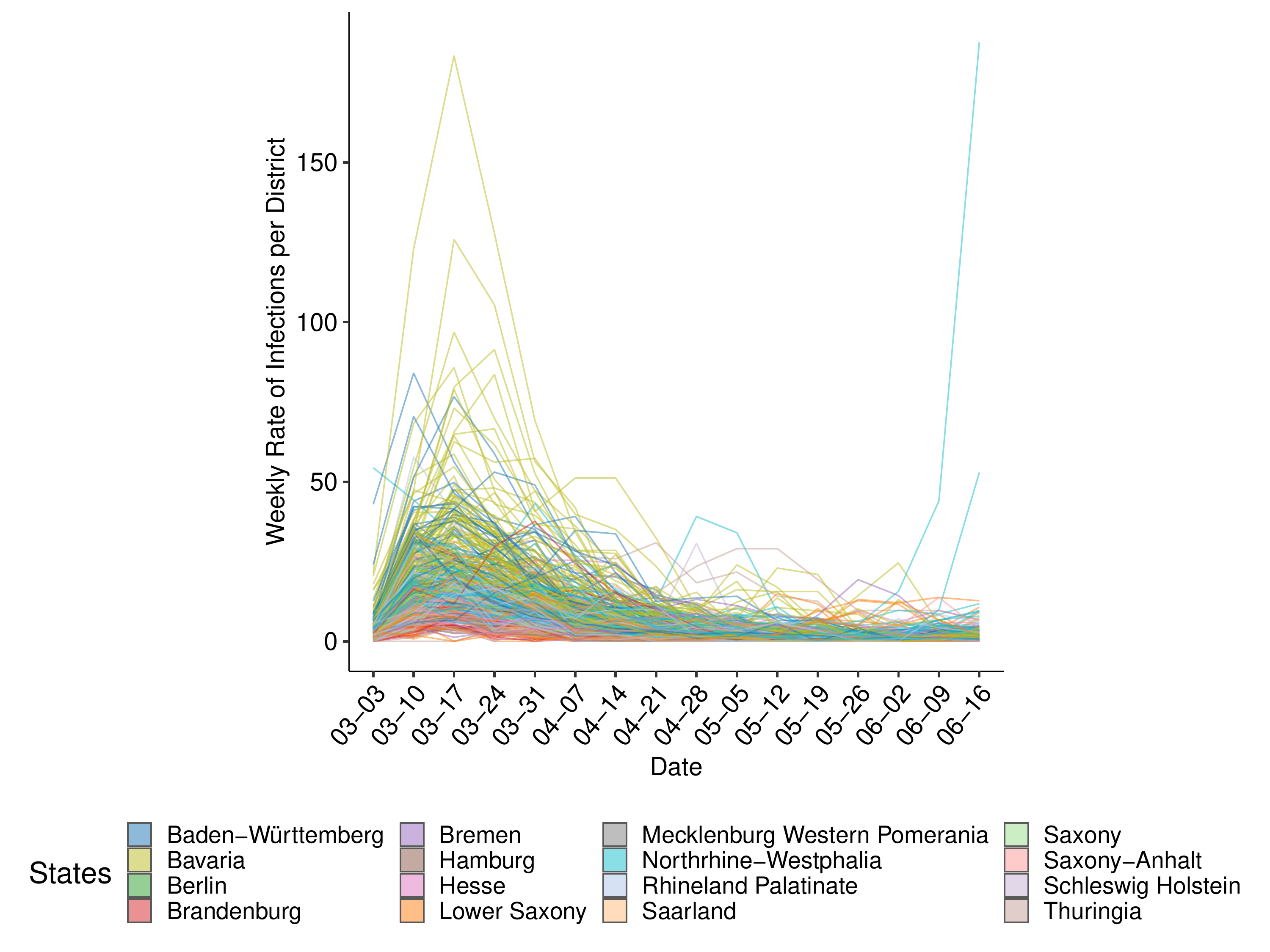}
		\caption{Observed Rate of Weekly Infections for each federal district. The colour of the lines indicate the federal state in which each district is located and the dates (mm:dd) are the first day of the corresponding week.}
		\label{fig:target}
	\end{figure}
	
	Our application's outcome of interest is the ratio of COVID-19 infections in a federal administrative district (NUTS-3 level), which we define as the quotient of the number of COVID-19 infections over the corresponding population size. In Germany, there are $n= 401$ federal administrative districts (a complete list is given by the \href{https://www.destatis.de/DE/Themen/Laender-Regionen/Regionales/Gemeindeverzeichnis/Administrativ/04-kreise.html}{German Federal Statistical Office}).  At a higher hierarchical level, each federal district also belongs to a  federal state (NUTS-1 level). In most figures, e.g., Figure \ref{fig:gini}, we colour-code the district-specific time series according to this allocation. If we refer to a specific district in the text, we generally specify the corresponding federal state in brackets. 
	
	\textbf{Infection count:}  The \href{https://hub.arcgis.com/datasets/dd4580c810204019a7b8eb3e0b329dd6_0/}{Robert-Koch-Institute} provides timely data on the daily number of COVID-19 infections in Germany for each federal district. We limit the present analysis to individuals between 15 and 59 years old due to the age structure in the Facebook population. Besides, the given surveillance counts are stratified by age group (15-35 and 36-59) and gender. For each entry, dates of symptom onset and reporting are given, although the onset date is partially missing. Our principal analysis is based on the disease onset date since it ensures more valid information on the infection incidence \citep{Guenther2020}. Imputation of the missing values is required (we present our method in Section \ref{sec:imput}). By $y_{i,g,t}$ we denote the observed (and partially imputed) counts of new onsets within district $i$, age/gender-group $g$ and week $t$. For completeness, we define with $x_{i,g}$ the corresponding indicator for the age/gender group. 
	
	\textbf{Population:} We obtained district-,age- and gender-specific population data from the \href{https://www.destatis.de/EN/Themes/Society-Environment/Population/Current-Population/_node.html}{German Federal Statistical Office}. To guarantee a consistent definition of age-groups, we categorised the data according to the two primary age groups according to which the infection data are reported, namely people between 15-35 and 36-59 years old. The corresponding time-constant covariate is denoted for age/gender-group $g$ in district $i$ by $x_{i,g,pop}$.  
	
	The observed rates per 10.000 inhabitants $\tilde{y}_{i,g,t} = \frac{10.000 y_{i,g,t}}{x_{i,g}}$  are visualised in Figure \ref{fig:target} colour-coded according to the different states. For each week, we plot the rate of disease onsets that we partially impute in case of missingness, as described in detail in the next section. Once the first peak of infections could be overcome, the cases in the aftermath are increasingly attributed to local outbreaks. Two districts, namely Guetersloh and Warendorf (North Rhine-Westphalia), experience a local outbreak in a meat factory during the last weeks of the observational period \citep{Kottasova}. This local outbreak encompasses 48$\%$ of all infections with disease onset in the week starting on the 16th of June.

	\subsection{Data on social activity during COVID-19}
	
	\label{sec:social_activity}
	\begin{figure}[t!]
		\centering
		\includegraphics[width=0.9\linewidth]{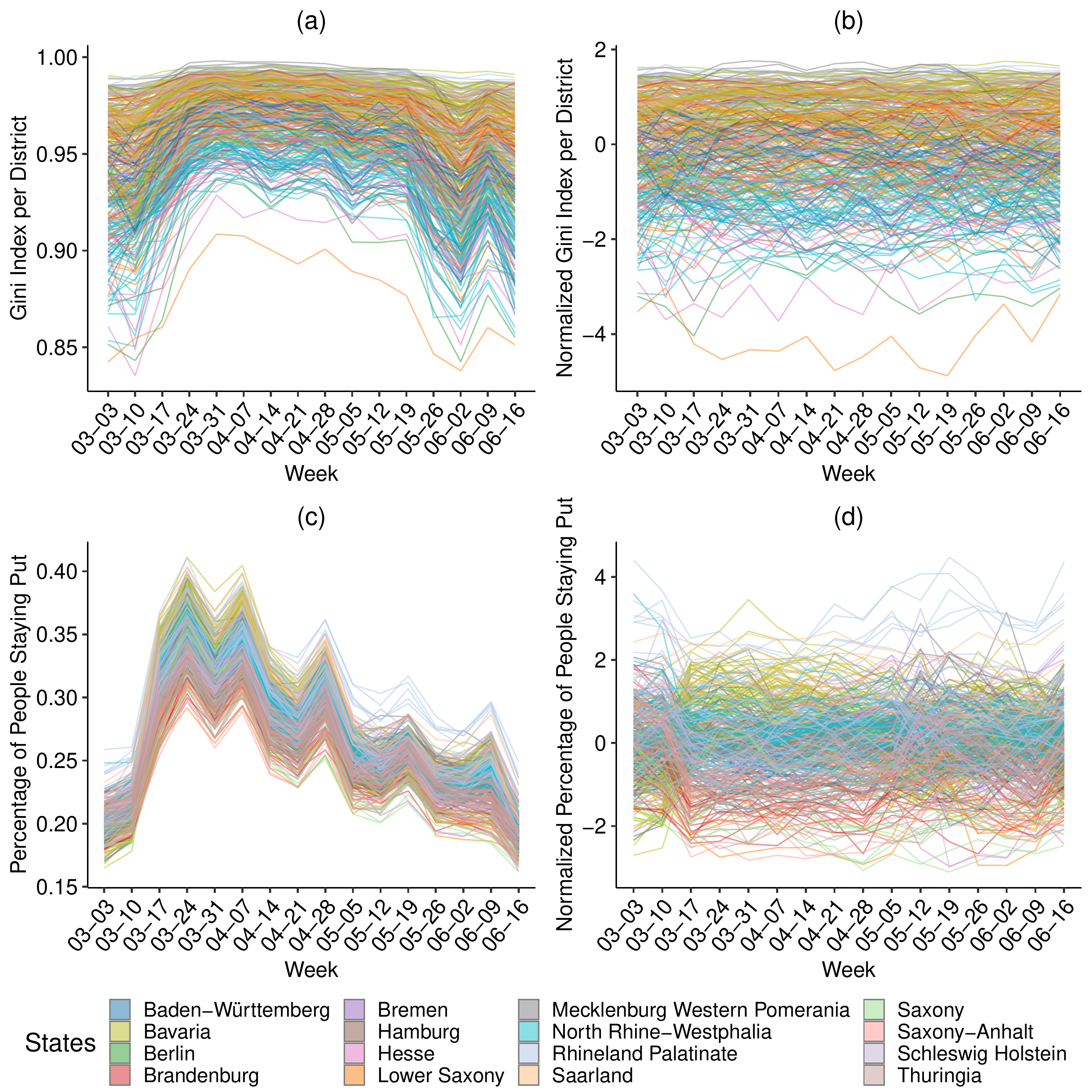}
		\caption{(a): Gini indices for each district over time. (b): Standardised Gini indices for each district over time. (c): Percentages of people staying put for each district over time. (d): Standardised percentages of the people staying put.  The colour of the lines indicate the state in which each district is located and the dates (mm:dd) are the first day of the corresponding week.} \label{fig:gini}
	\end{figure}

	All data related to social activities during the COVID-19 pandemic are generated from approximately 10 million Facebook users in Germany, who enabled geolocation features in the Facebook app on their mobile devices. To abide by the privacy policies, the observations are anonymised through aggregation onto tile bing polygons, censoring if we observed not enough users in the spatial region, as well as randomisation using additional noise and spatial smoothing \citep{Maas}. We aggregate the polygons to the same spatial units for our application on which we have the infection data. We propose the following measures describing social interaction and mobility. All measurements are taken weekly, where we use simple averaging for quantities available at a more granular temporal resolution.

	\textbf{Co-location:} Co-location in week $t$ is measured by the probability $p_{ij,t}$ of a random person from district $i$ to be located in the same $0.6km\times 0.6km$ square as another random person from district $j$ \citep{Iyer}. These probabilities are then used to construct a district-wise quantity for the concentration of meeting patterns using the Gini index, which is given by:
	\begin{align*}
	x_{i,t,gini} = \frac{\sum_{m, l \neq i} |p_{im,t} -p_{il,t} |}{2 (n-1) \sum_{j\neq i} p_{ij,t} }
	\end{align*}
	If we were to observe the maximal value of 1 in $x_{i,t,gini}$, all people within federal district $i$ would only meet people (i.e. Facebook users) from only one further district. This behaviour is exemplary of extremely restricted mobility. Conversely, a lower value heuristically indicates dispersed meeting patterns. Due to this intuitive interpretation, we opt for the Gini index as a measure of concentration. The Gini indices' temporal paths for the 401 districts in Germany are depicted in Figure \ref{fig:gini} (a). Overall, the meeting patterns become more concentrated on a few other districts as the crisis evolves. This behaviour contrasts rather dispersed practices before the pandemic. An upward trend is visible until the nationwide lockdown on 22nd of March, 2020\footnote{In Bavaria the lockdown started already on the 16th of March, 2020.}. Thereupon, meeting patterns continue to be overall condensed, although the indices slowly decline. To enable a meaningful comparison between the respective estimates in the regression setting of Section \ref{sec:model}, we standardise the Gini indices per week. The standardised covariate $\tilde{x}_{i,t,gini}$ is shown in Figure \ref{fig:gini} (b) and given by 
	\begin{align}
	\tilde{x}_{i,t,gini} = \frac{x_{i,t,gini} - \hat{\mu}_{t,gini}}{ \hat{\sigma}_{t,gini}},
	\label{eq:standardisation} 
	\end{align}
	where $\hat{\mu}_{t,gini} = \frac{1}{n} \sum_{j = 1}^n x_{j,t,gini} $ and $\hat{\sigma}_{t,gini} = \sqrt{\frac{1}{n-1} \sum_{j = 1}^n (x_{j,t,gini} - \hat{\mu}_{t,gini})^2 }$.
	
	\textbf{Percentage staying put:}  Besides the relative attribution of co-location probabilities to other districts, we investigate a measure that expresses how people (Facebook users)  comply with social distancing. We quantify this concept by the covariate $x_{i,t,sp}$, which is defined as the average percentage of people in district $i$ staying put during week $t$. Respective data were collected using geolocation traces of mobile devices and users are defined to be staying put, if they are only observed in one  $0.6km\times 0.6km$ square throughout a day \citep{Facebook}.  In Figure \ref{fig:gini} (c) clear break-points are visible, giving evidence of the temporary lockdown that started between the 17th and 24th of March. During the following weeks, the observed values gradually level off around pre-lockdown values. We also observe some peaks in the weeks starting on the 7th and 28th of April, which we traced back to the different mobility behaviour during national holidays, namely \textsl{Good Friday} on the 10th of April and \textsl{Labour Day} on the 1st of May 2020. 
	
	Similarly to the treatment of the Gini index, we standardise the percentages in the regression setting. While the visual impression from \ref{fig:gini} (c) insinuates that the dynamics of people staying put are similar between districts, the standardised paths, given in Figure \ref{fig:gini} (d), reveal local differences between them. For instance, the early look-down in Bavaria resulted in a substantial relative increase of the respective districts between the 10th and 17th of March, see the yellow-green lines.

	\begin{figure}[t!]
		\centering
		\includegraphics[width=0.9\linewidth]{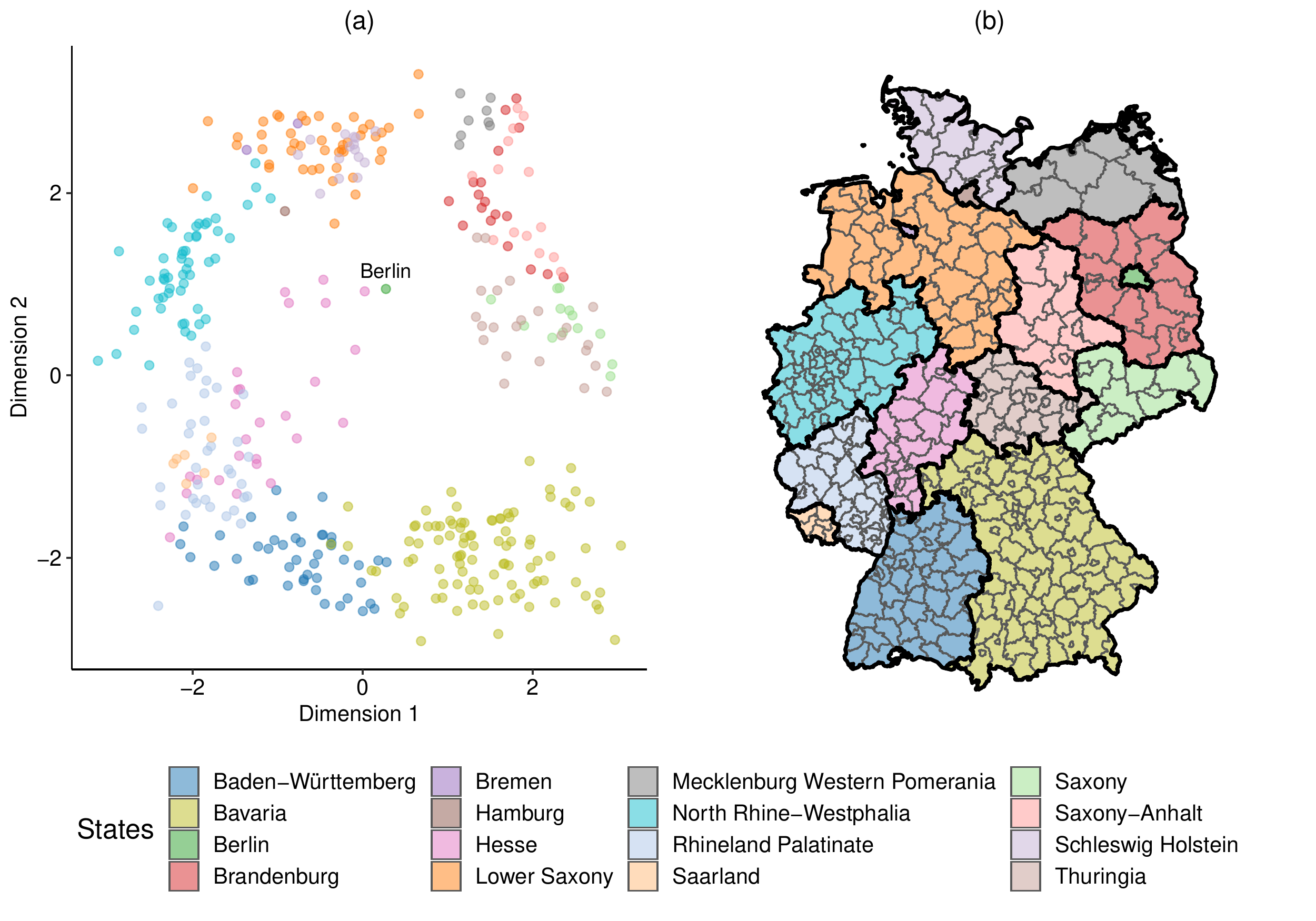}
		\caption{(a): Coordinates representing the friendship distances. The colour of the points indicates the state in which each district is located. (b): Map representing the colour legend. The thick black lines represent borders between federal states, while the thinner grey borders separate federal districts.}
		\label{fig:mds_real}
	\end{figure}
	
	\textbf{Friendship distance:} Spatial distance is found to be strongly associated with the spread between regions \citep{Kang2020}. Beyond the geographic proximity,  \citet{Cho2011} argued that friendship ties explain specifically long-distance mobility. This type of travel is fundamental for understanding the early spread of the pandemic \citep{Chinazzi2020}.  To accommodate this possible line of infection, we include a measure for the strength of friendship ties between the districts of Germany. More precisely, we employ the social connectedness index proposed by \citet{Bailey2018a}, which is based on an anonymised snapshot of all active Facebook users and their friendship networks from April 2020. For the administrative district $i$ and $j$ the time-invariant measure $x_{ij,soc}$ is given by: 
	\begin{align}
	x_{ij,soc}= \frac{\# \lbrace \text{Friendship Ties between users in district $i$ and $j$} \rbrace}{\# \lbrace \text{Users in district $i$} \rbrace \# \lbrace \text{Users in district $j$} \rbrace} ~ \forall~  i,j \in \lbrace 1, \ldots, n \rbrace. \label{eq:soc}
	\end{align}
	In a note, \citet{Kuchler2020} uncover high correlations between the social connectedness indices and the spread of COVID-19. This index is further processed to provide a spatial allocation based on social instead of Euclidean distances. To do so, we first transform social connectedness to social distance $d_{soc}$ by taking the reciprocal of connectedness, i.e., $d_{ij,soc} = \frac{1}{x_{ij,soc}}$. Consecutively, we process these distances to coordinates using \textsl{multidimensional scaling} \citep{Cox2000}. In our application, this procedure's result is a two-dimensional representation of each district in the network defined through \eqref{eq:soc} that is only identifiable up to the scale and rotation. Using \textsl{Procrustes analysis}, we map the rotation of the inferred coordinates in the friendship space to be most similar to the geographic coordinates \citep{Chen2008}. The outcome of the algorithm for each district $i$ is denoted by $x_{i,soc}$ and gives the geo-coordinates in the friendship space as shown in Figure \ref{fig:mds_real}. Robust connectivity within federal states and neighbouring districts are visible in the friendship coordinates. We also observe that the capital, Berlin, is situated in the very centre, reflecting its unique and highly connected position. One can also detect a persisting corridor between districts located in former East- and West-Germany. Next to the social coordinates, we incorporated each district's geographic coordinates $x_{i,coord}$, i.e., the longitude and latitude of each districts centroid, in our application.  Technical details on both procedures are given in Annex \ref{sec:mds}.
	
	
		\begin{figure}[t!]
		\centering
		\includegraphics[width=0.55\linewidth, page =1]{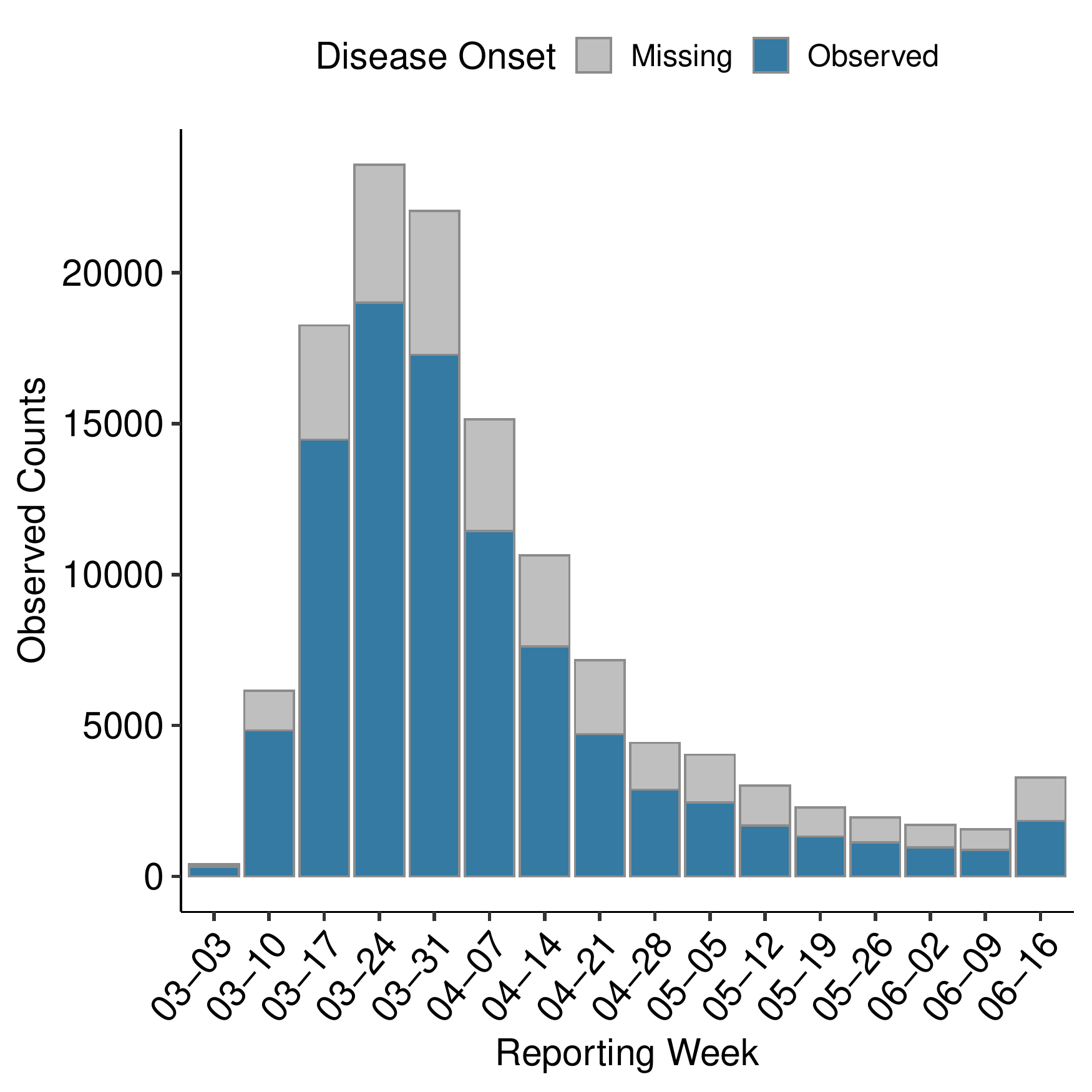}
		\caption{Count of missing and observed disease onset dates per reporting week.}
		\label{fig:impute}
	\end{figure}  
	\newpage
	\section{Modelling}
	\label{sec:model}
	
	We start by proposing a model to impute missing dates of the disease onset. Subsequently, these partially imputed infection data are modelled with a Negative Binomial regression.   
	
	\subsection{Imputation model}
	\label{sec:imput}

	We can see in Figure \ref{fig:impute} that approximately 30 $\%$ of the onset dates are missing. To still make use of all available information, we propose to impute missing disease onset dates under the assumption of missingness at random. This allows for unbiased findings which are not guaranteed when using a complete case analysis \citep{Little2002}. In particular, we leverage the fact that the chronologically later reporting date is available for all cases. Thereby, the problem of imputing the date of disease onset for a single infection is reduced to imputing the time between onset of disease and its reporting through a positive test, which we call test delay. Following \citet{Guenther2020}, we use the subset of all data without any missing disease onset dates to fit a distributional regression model for this test delay. In the next step, we predict all distributional parameters under this model for all cases with a missing disease onset date and sample the missing onset date.         
	
	To fit the imputation model, we first disaggregate the given surveillance counts to the patient level. For each complete case $l$, the data includes the age/gender group indicator ($x_{l,g}$),  an indicator whether the reporting date was during a weekend ($x_{l, weekend}$), and the state ($x_{l, state}$) and district ($x_{l, district}$) where it was observed. Regarding the temporal information of each infection, we are given the date of disease onset ($t_{l,o}$) and its reporting ($t_{l,r}$). For complete-case data, the test delay is then given by $d_{l} = t_{l,r} -t_{l,o}$. As regressors in the imputation model we include dummy covariates $x_l = (x_{l,g}, x_{l,weekend},x_{l, state},x_{l, district})$ and the metric covariate $t_{l,r}$ itself to account for changing testing strategies, e.g., during the early spread the test capacities were limited and patients needed to wait longer for a test to be conducted. We assume that $d_l$ is a realisation of random variable $D_l$, which follows a Negative Binomial model: 
	\begin{align}
	D_{l} \mid x_{l},t_{l,r} \sim NB\Big(\mu_{l} = \exp\big\lbrace\theta_\mu^\top x_{l} + f_\mu(t_{l,r})\big\rbrace,\sigma_{l} =  \exp\lbrace \theta_\sigma^\top x_{l} + f_\sigma(t_{l,r})\big \rbrace \Big), 
	\label{eq:negbin}
	\end{align} 
	where $\mathbb{E}(D_{l} \mid x_{l}) = \mu_{l}$ and $\text{Var}(D_{l} \mid x_{l}) = \mu_{l} + \sigma_l \mu_{l}^2$ holds.
	A discrete-valued distribution appears most suitable since the patient-level data are available daily, making the test delay inherently discrete. As indicated in \eqref{eq:negbin}, we model the location and scale parameters of the distribution by separate linear predictors. Note that the linear predictors are defined by $\eta_\mu =\theta_\mu^\top x_{l} + f_\mu(t_{l,r})$ and $\eta_{\sigma} = \theta_\sigma^\top x_{l} + f_\sigma(t_{l,r})$ for the corresponding distributional parameters and that the linearity only refers to linearity in the coefficients not in the covariates. Therefore, the model lies within the family of generalised additive models for location, scale and shape \citep{Rigby2005}. While all components of $x_l$  have a log-linear effect, we parametrise the trend effect of the reporting date $t_{l,r}$ by nonlinear penalised splines (see \citealp{eilers1996} for details). The district-specific effects are assumed to be Gaussian.  After having obtained the estimates, we calculate $\hat{\mu}_{\tilde{l}} = \exp\big\lbrace\hat\theta_\mu^\top x_{\tilde{l}} + \hat f_\mu(t_{\tilde{l},r})\big\rbrace$ and $\hat{\sigma}_{\tilde{l}} = \exp\lbrace \hat\theta_\sigma^\top x_{\tilde{l} } + \hat f_\sigma(t_{\tilde{l} ,r})\big \rbrace$ for all observations $\tilde{l}$ with missing disease onset. We can now simulate $d_{\tilde{l}}$ from \eqref{eq:negbin} to acquire a full dataset by setting $t_{\tilde{l},o} = t_{\tilde{l},r} - d_{\tilde{l}}$.  Through aggregation from the daily patient-level data to the infection counts per district $i$ and age/gender group $g$ with disease onset in week $t$, denoted by $y_{i,g,t}$, we build a single partially imputed dataset. This procedure is repeated $K$ times to represent the uncertainty associated with the missing information of all disease onsets. 
	
	\subsection{Infection model}
	\label{sec:infect}
	%
	
	To model the rate of infections with partially imputed data, we apply a negative binomial `\textsl{observation-driven}' model for count data including the population as an offset term \citep{Cox1981}. By doing so, we assume  
	\begin{align}
	Y_{i,g,t} \mid x_{i,g,t-1}, a_i, b_i \sim NB\big(\mu_{i,g,t}, \sigma	\big), \forall ~ i \in \lbrace 1, ..., 401\rbrace,  g\in \mathcal{G}, \text{and } t = 2, ..., T,
	\label{eq:nb}
	\end{align}
	where $x_{i,g,u} = (u,  x_{i,g}, x_{i,g,pop}, \tilde{x}_{i,u,gini}, \tilde{x}_{i,u, sp},x_{i,coord}, x_{i,soc},\tilde{y}_{i,g,t-1})$ are the covariates at arbitrary week $u$ specified in Section \ref{sec:data} and $\mathcal{G}$ denotes the set of age/gender groups used from the data. Further, let $T$ be the final week of data we use in the analysis. We assume in \eqref{eq:nb} that the random variable $Y_{i,g,t}$ follows a negative binomial distribution conditional on  $x_{i,g,t-1}, a_i$ and $b_i$ to compensate overdispersion in the observed counts. 
	
	Aligned with models for the spread of infectious diseases \citep{Held2005}, we decompose $\mu_{i,g,t}$ into  an endemic and epidemic component: 
	\begin{align}
	\mu_{i,g,t} = \exp\lbrace \nu_{i,g,t}^{END} + \nu_{i,g,t}^{EPI} \rbrace,
	\label{eq:end-epi}
	\end{align}
	where each part is parametrised as follows:
	\begin{align}
	\nu_{i,g,t}^{EPI} =&  \theta_{AR(1)} \log(\tilde{y}_{i,g,t-1} + c) 
	\label{eq:epi-param} \\
	\nu_{i,g,t}^{END} =& \theta_t + \theta_{gen} \mathbb{I} (x_{i,gen} = \text{``Male''})  + \theta_{age} \mathbb{I}  (x_{i,age} = \text{``36-59''})  + \label{eq:end-param}\\
	&+\theta_{age:gen}  \mathbb{I} (x_{i,gen} = \text{``Male''}) \cdot \mathbb{I}  (x_{i,age} = \text{``36-59''}) 
	+ \theta_{t,gini} {x}_{i,t-1,gini} \nonumber\\&+\theta_{t,sp} {x}_{i,t-1,sp}+ f_{coord}(x_{i,coord}) + f_{soc}(x_{i,soc})  + a_i + b_i \mathbb{I} (t = T) + \log(x_{i,g,pop}) \nonumber.
	\end{align} 
	We include a first-order autoregressive term of this rate, since path dependencies and self-exciting behaviour are common with infectious diseases and should therefore be accounted for \citep{Held2005}.  In addition, we transform the respective term by $h(x) =\log(x + c)$ to bypass problems with absorbing states of the implied counting process when $\tilde y_{i,g,t-1} = 0$. The value $c \in (0,1]$ is estimated from the data. More general types of these autoregressive models are proposed by \citet{Zeger1988}. 
	
	As is evident from \eqref{eq:end-epi}, we constitute that both the epidemic and endemic components have a multiplicative effect on the observed infection rates. As an alternative, \citet{Held2005} replace the log link by an identity link, although \citet{Fokianos2020} argue for the logarithmic link implied in \eqref{eq:end-epi} if additional covariates are available. They further derive theoretical properties, such as ergodicity, in the case of Poisson-distributed target variables under the condition $\theta_{AR(1)}<1$. 
	
	\textbf{Time-varying effects:}   For the endemic part \eqref{eq:end-param}, the temporal trend is reflected by piecewise constant fixed effects separately for each week, $\theta_t$. By means of group-specific covariates we control for gender- and age-related effects and their interaction,  $\theta_{gen}, \theta_{age}$, and $\theta_{age:gen}$ \citep{Walter2020}.  The principal covariates, Gini Index and Percentage Staying Put, are modelled by piecewise constants in each week for maximal flexibility. To account for the stylised fact, that the incubation period, i.e., the time between being infected and symptom onset, for COVID-19 is around 5 days \citep{Li2020}, we lag the information on Gini Index and Percentage Staying Put by one week as indicated in \eqref{eq:end-param}.  
	
	\textbf{Isotropic smooth effects:} The bivariate functions $f_{coord}(x_{i,coord})$ and $f_{soc}(x_{i,soc})$ display the effects of geographic coordinates and social coordinates on the incidence rate. To properly incorporate $x_{i,coord}$ and $x_{i,soc}$ in our regression framework, we propose the usage of isotropic splines. These kind of flexible functions were proposed by \citet{Duchon1977} to model multiple covariates by a multivariate term as an alternative to anisotropic tensor products. Isotropic smoothers have the property of giving the identical predictions of the response under arbitrary rotation and reflection of the respective covariates \citep{wood2017}. This characteristic is commonly reasonable when working with geographic coordinates $x_{i,coord}$ and in accordance with the uniqueness of the multidimensional scaling results, thus also for $x_{i,soc}$. With respect to the form of the smooth terms, we follow \citet{wood2003} and use a low-rank approximation of the thin-plate splines introduced in \citet{Duchon1977}. To obtain a smooth fit, we impose a penalty that is controlled by $\tau_{soc}$ and $\tau_{coord}$ for the respective isotropic splines. 
	
	\textbf{Random effects:}   Because super spreader events such as carnival sessions \citep{Streeck2020} or local outbreaks in major slaughterhouses (\citealp{Dyal2020}) lead to unobserved heterogeneities, our model comprises two district-specific Gaussian random effects. The random effect $a_i$ governs long-term heterogeneities, while short-term dependencies, i.e., sudden locally confined outbreaks as visible in the last week of Figure \ref{fig:target}, are captured by $b_i$. We assume $a = (a_1, \ldots, a_n)^\top \sim N(0,I_n\tau_a^2)$ and $b = (b_1, \ldots, b_n)^\top \sim N(0,I_n\tau_b^2)$. Relying on the duality between semiparametric regression and random effects \citep{Ruppert2003a}, we can equivalently write the random effects as semiparametric terms. Hence we may replace $a_i$ and $b_i$ by $f_a(i) = (a_1, ..., a_n)^\top X_a$ and $f_b(i) = (b_1, ..., b_n)^\top X_b$, respectively, and introduce a ridge penalty for each coefficient vector. In this context, the design matrices $X_a$ and $X_b$ each consist of $n$ dummy variables indicating to which district a specific observation refers. As a result of this reformulation, we can estimate the additional parameters $\tau_a$ and $\tau_b$ as tuning parameters in semiparametric regression (see Annex \ref{sec:ann_estim} for further information).
	
	
	\textbf{Modelling rates via count regression:}  Effectively, we model the rate of infections by  including the term $\log(x_{i,g,pop})$ as an offset in \eqref{eq:end-param} since the infections rates  $\tilde{Y}_{i,g,t}$ relate to the counts through $Y_{i,g,t} =\tilde{Y}_{i,g,t} x_{i,g,pop}$  via (note the slight abuse of notation as we here do not regard the infection rate among 10.000 inhabitants but the percentage of people infected with a disease onset in a specific week). As a byproduct, we implicitly assume that the entire population is susceptible, which is reasonable when considering the low prevalence of COVID-19 in Germany during the first wave. However, the model is still applicable in the later stages of the pandemic by replacing this offset with the number of susceptible inhabitants in each region.  
	
	

	
	%
	
	
	\subsection{Estimation}
	\label{sec:estimation}
	
	At first, we propose an estimation procedure for the imputation model from Section \ref{sec:imput}. Given a partially imputed dataset, we specify how to get estimates for the infection model from Section \ref{sec:infect}. Finally, the multiple imputation scheme combining both approaches is presented. Generally, we carry out all computations conditional on the observations in $t = 1$, i.e., the week between the 3rd and 9th of March.


	\textbf{Imputation model:}  We get estimates for the imputation model through maximising the likelihood function resulting from \eqref{eq:negbin}. As mentioned in Section \ref{sec:infect}, we can rewrite all random effects as smooth terms and penalise the likelihood to obtain smooth functions. By repeatedly updating the estimators through a backfitting algorithm, we maximise this objective (see \citealp{Rigby2005} for details). This procedure is readily implemented in the software package \texttt{gamlss} \citep{gamlss.add}. 
	
	\textbf{Infection model:} The infection model is characterised by the parameters $c$ and $\theta$, relating to the log-transformation of the autoregressive component and all other parameters. Given a partially imputed dataset, we first consider $\theta$ to be a nuisance parameter and find $c$ via a profile likelihood approach.  Here the profile likelihood is given by
	\begin{align*}
	\mathcal{L}_{Profile}(c) = \underset{\theta}{\text{max }}\mathcal{L}(c,\theta) = \mathcal{L}(c,\hat{\theta}(c)), 
	\end{align*}
	where $\mathcal{L}(c,\theta)$ is the joint likelihood resulting from \eqref{eq:nb} and $\hat{\theta}(c)$ is the maximum likelihood estimator of $\theta$ for a fixed value of $c$. For any $c$ we can find $\hat{\theta}(c)$ by carrying out the estimation as explained in Annex \ref{sec:ann_estim}, hence it is straightforward to evaluate $\mathcal{L}_{Profile}(c)$. Building on this result, we use standard optimisation software, i.e., the $\mathtt{optimise}$ routine within the software environment $\mathtt{R}$ \citep{R},  to obtain $\hat{c} = 
	\text{arg} \underset{c}{\text{ max }} 	\mathcal{L}_{Profile}(c)$. In the consecutive step, we fix $c$ at $\hat{c}$ to get $\hat{\theta}$ again by following Annex \ref{sec:ann_estim}. 
	
	\textbf{Multiple imputation:} Since information on the onset of symptoms is missing for approximately $30\%$ of the cases, we proposed an imputation model in Section \ref{sec:imput} to generate $K$ partially imputed datasets. To correct the uncertainty quantification of the infection model for this multiple imputation procedure, we use the Rubin's rule. At first, we sample $K$ imputed datasets according to Section \ref{sec:imput}. Let $\hat{\vartheta}_{(k)} = (\hat{\theta}_{(k)}, \hat{c})$ be the resulting estimator from  the two-stage maximum profile likelihood procedure explained in the previous paragraph given the partially imputed dataset from the $k$th imputation step. By $\hat{V}_{(k)}$ we denote the corresponding variance estimate that results from Bayesian large sample properties \citep{wood2013}. We then average the coefficients over all $K$ iterations to obtain  $\hat{\vartheta}_{MI} = \frac{1}{K} \sum_{k = 1}^K \hat{\vartheta}_{(k)}$ and estimate its variance through: 
	\begin{align*}
	\hat{\text{Var}}(\hat{\vartheta}_{MI}) = \bar{V} + (1 + K^{-1}) \bar{B}, 
	\end{align*}
	where its components are given by
	\begin{align*}
	\bar{V} &= \frac{1}{K} \sum_{k = 1}^K \hat{V}_{(k)} \\
	\bar{B} &=  \frac{1}{K -1} \sum_{k = 1}^K \big(\hat{\vartheta}_{(k)} - \hat{\vartheta}_{MI}  \big) \big(\hat{\vartheta}_{(k)} - \hat{\vartheta}_{MI}  \big)^\top.
	\end{align*}
	In our application, setting $K = 20$ proved to be sufficient since the estimates of different imputed datasets varied only marginally. 
	\section{Results}
	\label{sec:results}
	
	

	We only report the findings of the infection model detailed in Section \ref{sec:infect}. A detailed analysis of the imputation model as well as a robustness check for the infection model can be found in the Supplementary Material.
	
	\subsection{Temporal effect}
	
	\begin{figure}[t!]
		\centering
		\includegraphics[width=0.45\linewidth, page =1]{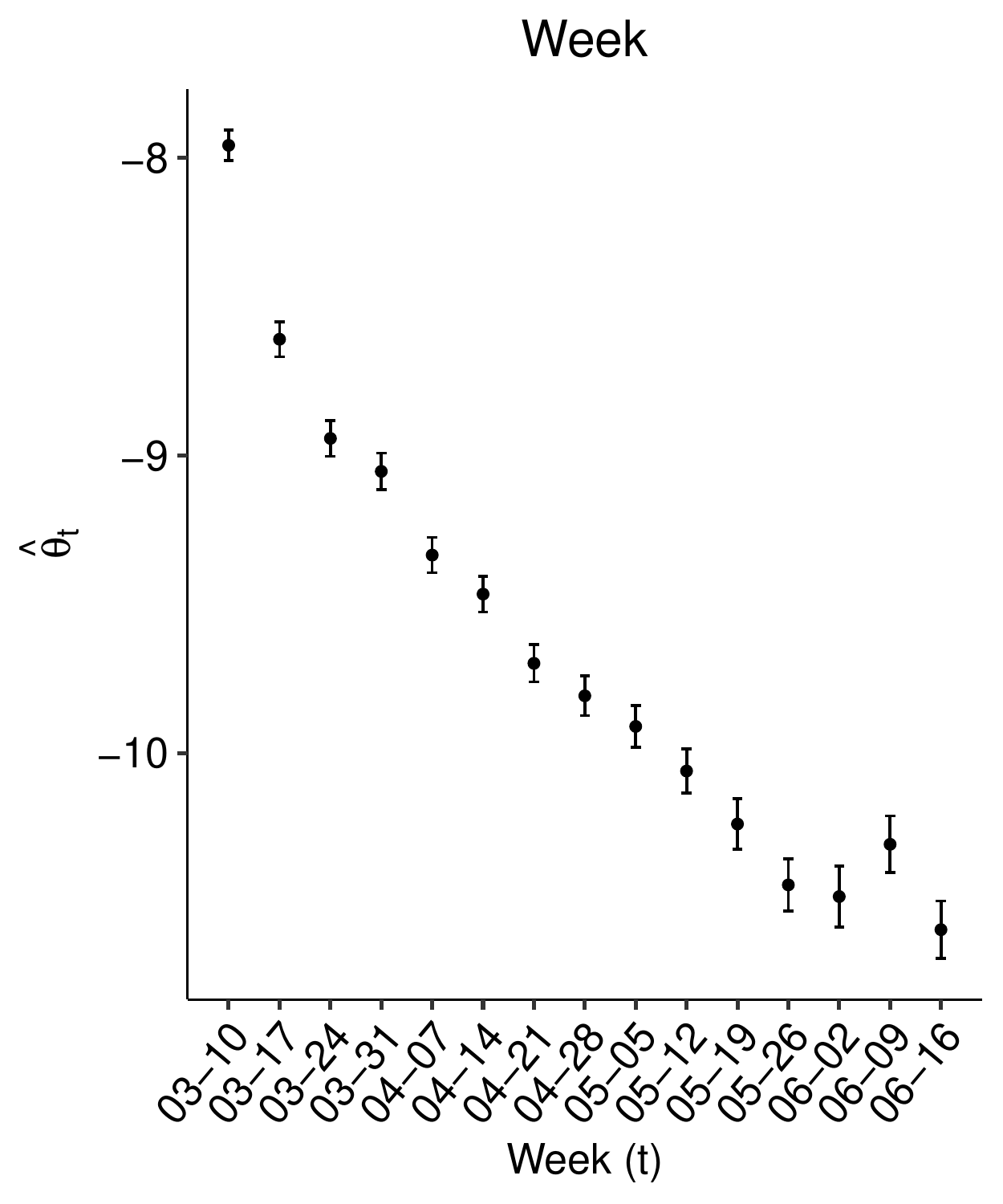}
		\caption{Estimate of temporal effect $\theta_t$. The $95\%$ confidence interval accompanies the estimates, and the shown dates (mm:dd)  on the x-axis are the first days of the corresponding weeks.}
		
		\label{fig:base2}
	\end{figure}
	To start, the estimate of $\theta_t$ is shown in Figure \ref{fig:base2}. The progression of the weekly estimates confirms generally decreasing infection rates over time. Due to the standardisation employed for the principal covariates in the analysis, the temporal trend can be interpreted as the log-transformed expected infection rate of female individuals aged between 15 and 35 in a district where the standardised Gini Index and Percentage Staying Put are zero. Since observing a zero in the standardised covariates translates to the mean observed values where we observed most information, the standard errors are also extremely narrow.

	\subsection{Sociodemographic and epidemic effects}
	\begin{table}[t!]
		\begin{center}
			\begin{tabular}{l | c c}
				Covariable & Estimate & exp$\lbrace$Estimate$\rbrace$\\
				&  (Standard Error) &  (Standard Error) \\
				\hline
				
				Male                    & $ -0.03$ &     0.97 \\
				& $(0.015)$ &   $(0.014)$   \\
				A35-A59            & $-0.031  $&  0.969   \\
				& $(0.014)$     &$(0.013)$ \\
				Male:A35-A59            & $-0.071$&   0.931  \\
				& $(0.02)$     &$( 0.017)$ \\
				$\log(\tilde{y}_{i,g,t-1} + c)$     &               $0.623$   & 1.865    \\
				& $(0.009)$   &   ( 0.031) \\
			\end{tabular}
			\caption{Estimates of linear time-constant effects.The reference group are female individuals aged between 15 and 35. By use of the delta rule we approximated the standard errors of the transformed coefficients in the third row. The value $c$ is estimated at 0.499 with a standard error of 0.027.}
			\label{table:coefficients}
		\end{center}
	\end{table}
	
	The linear time-constant estimates are given in Table \ref{table:coefficients} and exhibit in general a negative effect on male patients compared to female patients, $3\%$ in the younger and $ 9.6\%$ in the older age cohort\footnote{One can derive these percentages by computing the expected multiplicative change that results from alternating the prediction from one to another demographic group. For instance, $\exp\{\-0.03\} \approx 0.97$, which is equivalent to a $3\%$ decrease, is the multiplicative change ceteris paribus between females and males both aged between 15 and 35.}. According to its partial effect, we also predict that the older age group has a lower infection rate than the younger group encompassing individuals aged between 15 and 35, for men $9.7\%$ and women $3.1\%$. The autocorrelation coefficient $\theta_{AR(1)}$ expresses that one more infection among 10.000 inhabitants in a district during the past week almost doubles the predicted infections for the present week. This dominant finding confirms strong path dependencies in the data. In this context, we need to remark that the coefficients are partial effects that condition on all other covariates used in the model. Therefore, a positive coefficient of a dummy variable does not necessarily translate to the same finding in the raw numbers.
	
	\subsection{Mobility effects }

	\begin{figure}[t!]
		\centering
		\includegraphics[width=0.45\linewidth, page =1]{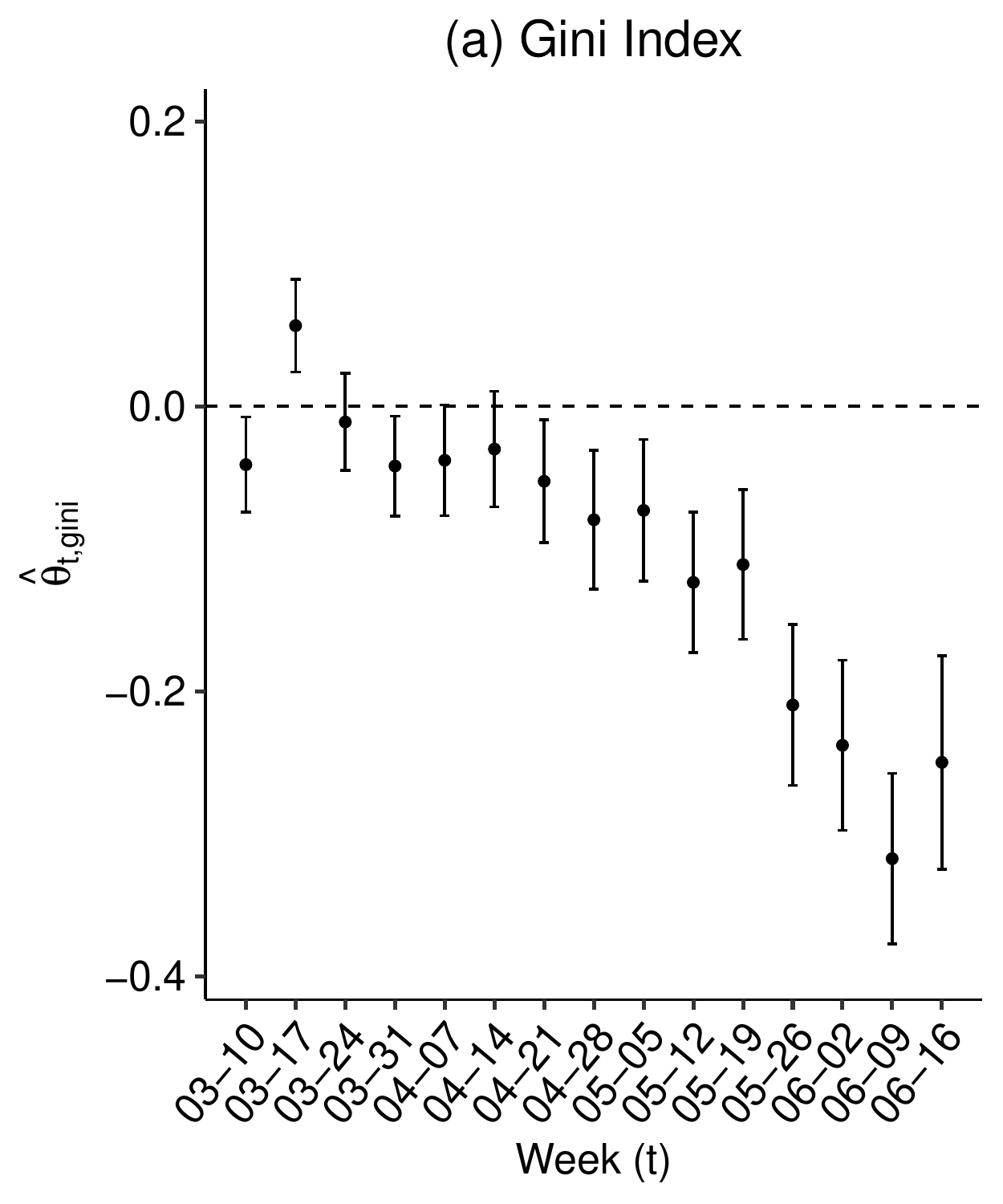}
		\includegraphics[width=0.45\linewidth, page =1]{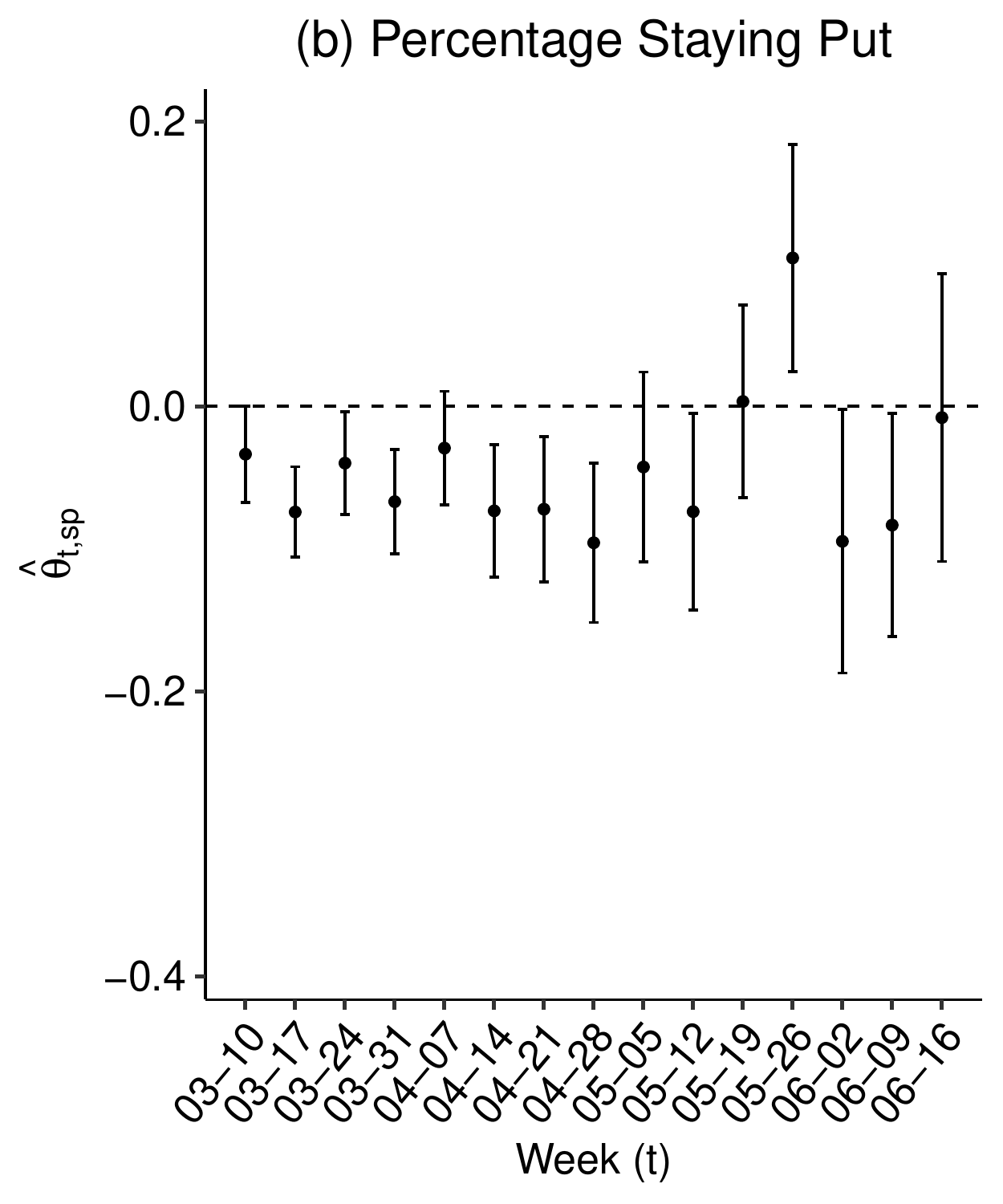}
		\caption{(a): Time-varying effects of the Gini index $\hat{\theta}_{t,gini}$. (b): Time-varying effects of the Percentage of People Staying Put $\hat{\theta}_{t,sp}$. The $95\%$ confidence interval accompanies the estimates, and the shown dates (mm:dd)  on the x-axis are the first days of the corresponding weeks.}
		\label{fig:estimates}
	\end{figure}

	The time-varying estimates regarding the relative mobility pattern are displayed in Figure \ref{fig:estimates}. Overall, the estimated effects of the measures proposed in Section \ref{sec:social_activity} on the rate of local COVID-19 infections are negative. In regards to relative importance, both variables rank similarly during the lockdown period that persists until early May. Subsequently, the Gini Index in a region gains weight, while the effect of People Staying Put becomes more volatile. The temporal changes of the respective estimates illustrate nonlinearities, which would not have been sufficiently captured by linear effects. 
	
	
	\textbf{Gini index of co-location}: Given all other covariates, Figure \ref{fig:estimates} (a)  suggests that inhabitants with meeting patterns that are centred around a few other districts entail reduced infection rates for a specific district. This tendency is only suspended in the week starting on March 17th during the early lockdown in Bavaria. The corresponding estimate is positive and significant. Right after the national lockdown on March 22nd, 2020, is ordered, the effect is not significantly different from zero for one week (03-24). The estimated effects remain low but negative until the German government introduces compulsory masks in public areas on April 22nd \citep{Mitze2020}. Thereupon, the effect has a clear downwards tendency.  Once policymakers slowly lift the lockdown measures, the estimate declines further until its maximum in the penultimate week of our observational period. This development may be viewed as evidence that a more focused attribution of co-location probabilities in a district becomes more crucial over time. 

	\textbf{Percentage staying put}: Suppose the percentage of inhabitants in a district staying put is large relative to the national tendency. In that case, we expect the incidence of infections throughout the lockdown period to be lower. We deduce this result from the largely negative estimates in Figure \ref{fig:estimates} (b) for the weeks between March 10th and May 12th.  Once the orders are relaxed, on the other hand,  the standard errors of the respective covariate become relatively large, and the effect vanishes in the final week of the study. A possible explanation for this phenomenon is that when daily infections decline, most diseases are related to local outbreaks (as already mentioned in Section \ref{sec:data}). These breakouts, in turn, can not be associated with the percentage of people staying put. One exception to this finding is the estimate in the week starting on May 26th, where we encounter a significant positive effect.
	\subsection{Spatial and social connectedness effects}
	
	\begin{figure}[t!]
		\centering \hspace{2cm}
		\includegraphics[width=0.69\linewidth, page =1]{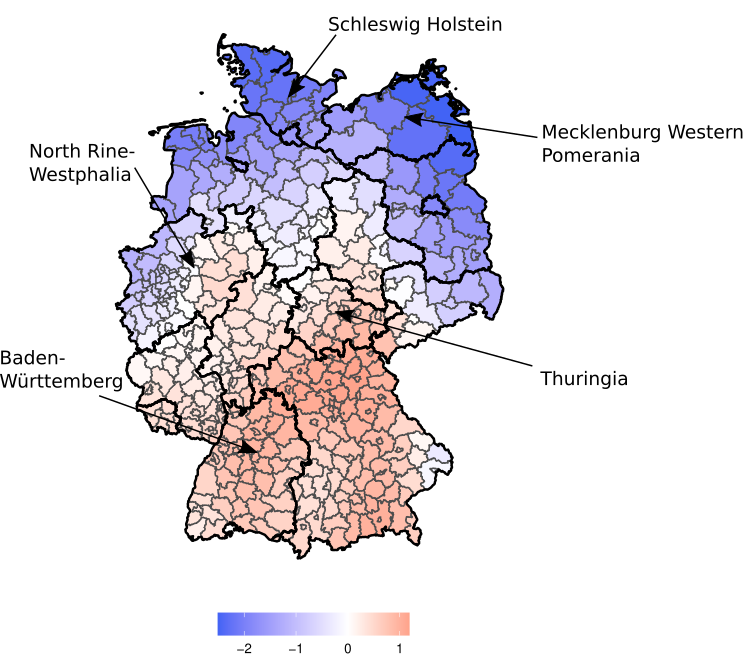}
		\caption{Estimated smooth spatial effect $f_{coord}$. The thick black lines represent borders between federal states, while the thinner grey borders separate federal districts. Through arrows, we highlight selected states mentioned in the text.}
		\label{fig:random}
	\end{figure}
	\begin{figure}[t!]
		\centering
		\includegraphics[width=0.9\linewidth]{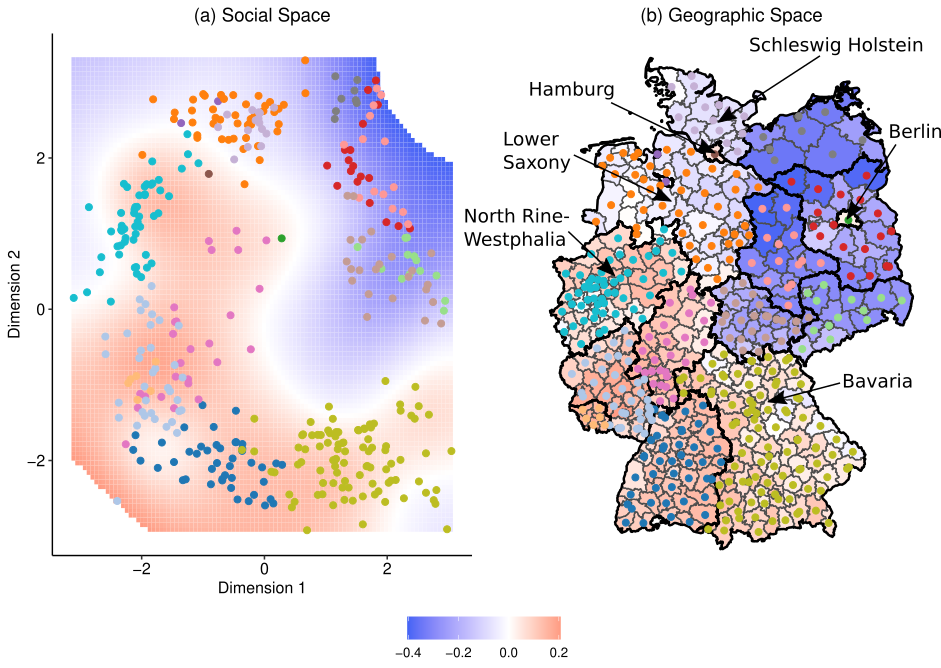}
		\caption{(a): Coordinates of the districts in the friendship space with the smooth partial effect of $f_{soc}$ in the background. We only show the predictions in the range of observed values. (b):  Coordinates of the districts in the geometric space with the smooth partial effect of $f_{soc}$ again shown in the background for each district. The thick black lines represent borders between federal states, while the thinner grey borders separate federal districts. Through arrows, we highlight selected states mentioned in the text.}
		\label{fig:mds}
	\end{figure}
	
	In our model specification, we incorporate the friendship coordinates and geographic coordinates as two spatial effects. In combination with the two unstructured latent variables, we can disentangle separate influences on the local infection rates of spatial and friendship proximity as well as short- and long-term district-specific deviations from it.

	\textbf{Spatial effects}: Let us start with the smooth spatial effect in Figure \ref{fig:random}. Overall, the geographic effects within federal states, indicated by the black borders in Figure \ref{fig:random}, are mostly heterogeneous. 
	To give some examples, an almost uniformly augmented risk of infections is estimated in  Baden-Württemberg and Thuringia. At the same time, we remark a negative spatial effect in Germany's northern districts, i.e., Schleswig Holstein and Mecklenburg Western Pomerania. On the other hand, the fit for districts in North Rhine-Westphalia varies between positive, negative, and no effect.

	We visualise the result of the friendship coordinates in two manners. One may plot the smooth bivariate function in the friendship space, Figure \ref{fig:mds} (a), or map the smooth fit on the geographic space, Figure \ref{fig:mds} (b). The re-mapping allows for sharp edges in the geographic coordinates. Broadly, the fit differentiates between districts allocated in former East-Germany (corresponding to MDS coordinates located in the first quadrant in Figure \ref{fig:mds} (a)) and former West-Germany. We observe that the predicted infections are ceteris paribus lower if a district is situated in former East-Germany. Districts allocated in the second and fourth quadrant of Figure \ref{fig:mds} (a) (mainly including districts from the states Bavaria, North Rhine-Westphalia and parts of Lower Saxony) are negatively affected by social proximity. Figure \ref{fig:mds} (b) demonstrates how the partial effects sometimes change abruptly between large cities and neighbouring districts. For instance, Berlin's central position is unrelated to the infection rates compared to the negative effect evaluated in Brandenburg. We observe a similar phenomenon for Hamburg when contrasting its partial effect with surrounding districts in Schleswig Holstein and Lower Saxony.

	
	\begin{figure}[t!]
		\centering
		\includegraphics[width=0.9\linewidth]{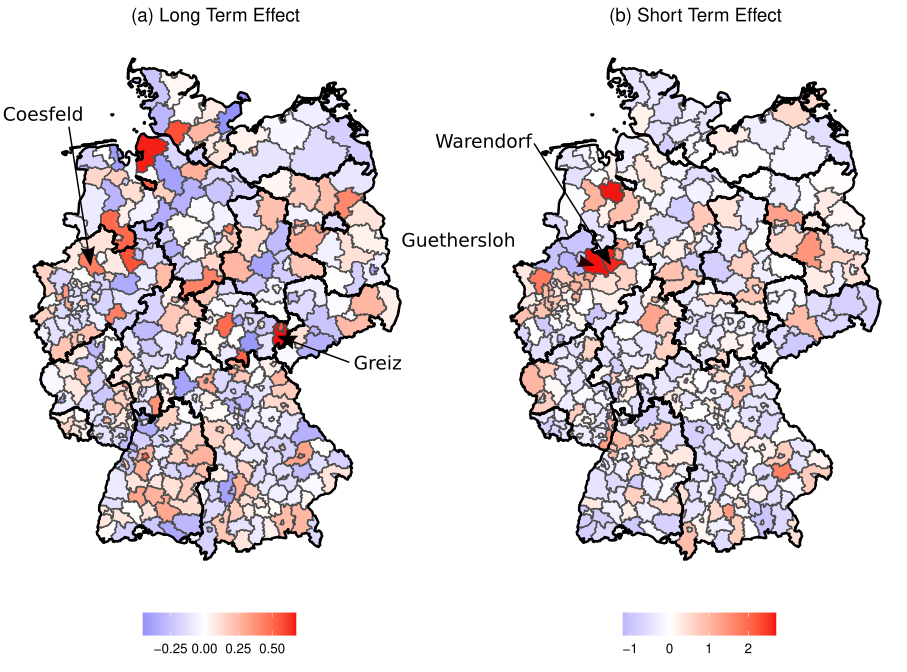}
		\caption{(a): Maximum posterior modes of the long-term random effects $a_i$. (b) Maximum posterior modes of the short-term random effects $b_i$. The thick black lines represent borders between federal states, while the thinner grey borders separate federal districts. Through arrows, we highlight selected districts mentioned in the text.}
		\label{fig:random_effects}
	\end{figure}

	\textbf{Unobserved heterogeneity effect}: In Figure \ref{fig:random_effects} the posterior modes of both random effects evince strong heterogeneities between districts  and underpin local differences in the spread of COVID-19. Noticeable estimates of the long-term random effects, Figure \ref{fig:random_effects} (a), reflect early outbreaks in the districts Greiz (Thuringia) and Coesfeld (North Rhine-Westphalia). Some estimates may also be related to heterogeneous testing practices between the districts. 
	
	We can trace back most high estimates of the short-term random effect to locally confined outbreaks, for instance, Guethersloh and Warendorf (North Rhine-Westphalia). As already stated in Section \ref{sec:data} the proportion of infections attributed to these local events rises once the general level of new cases declines. This result is supported by the different scales of the two types of random effects and apparent in the estimates $\hat{\tau}_a = 0.2 < \hat{\tau}_b =  0.585$. Therefore, the posterior modes of the short-term effects exhibit higher variances and are larger in absolute terms than the long-term effects.     
	
	
	\subsection{Model assessment}
	
	\begin{table}[t!]
		\center
		\begin{tabular}{l|c c}
			Model Description & cAIC (Model) & $\Delta$cAIC (Model)\\
			\hline
			Our Model & 86694 & --  \\
			With State Effect &  86694.79 &  0.790  \\
			Without Geographic Distance  & 86699.42 &  5.422  \\
			Without Friendship Distance & 86701.26 &  7.262 \\
			Without Age:Gender Interaction  &  86707.34 & 13.336 \\
			Without Percentage Staying Put & 86732.46 &  38.461 \\
			Without Gini Index& 86974.32 &   280.319  \\
			Without Facebook Covariates &  87033.87 & 339.867\\
			Without Long-Term Effect &  87452.62 & 758.620\\
			Without Short-Term Effect & 87900.03  & 1206.034\\
			Without Long- and Short-Term Effect & 88624.38 & 1930.382\\
			
		\end{tabular}
		\caption{Alternative model specifications with resulting corrected AIC value and change in corrected AIC value when compared to our model from Section \ref{sec:model}.}
		\label{tbl:aic}
	\end{table}
	
	We compare various alternative model specifications to check the robustness of our conclusions. In particular, we estimate separate models, adding dummy covariates for each state and leaving out one of the spatial terms, the Gini index, the Percentage of People Staying Put, all Facebook-related covariates, and random effects. 
	For this endeavour, we utilise the corrected Akaike Information Criterion (cAIC) introduced by \citet{wood2016} since the effective degrees of freedom need to adjusted for the additionally estimated variance components if random effects are included (we average the respective values over all imputed datasets). The results in Table \ref{tbl:aic} support the appropriateness of our final model since the corresponding cAIV value is the lowest. Besides, the change in the cAIC value to the model \eqref{eq:nb}, denoted by $\Delta$cAIC, permits an evaluation of the variable importance of each eliminated covariate. We can conclude from Table \ref{tbl:aic} that the exclusion of the Gini Index induces the highest loss in cAIC value. Concerning the different types of distances, the friendship distance is more important than the geographic distance.  
	
	\begin{figure}[!t]
		\centering
		\includegraphics[width=0.45\linewidth]{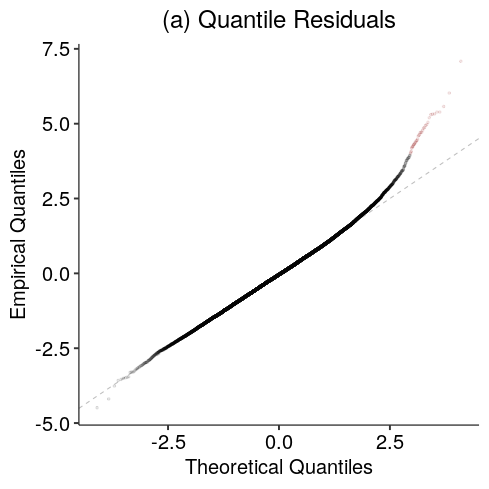}
		\includegraphics[width=0.45\linewidth, page =1]{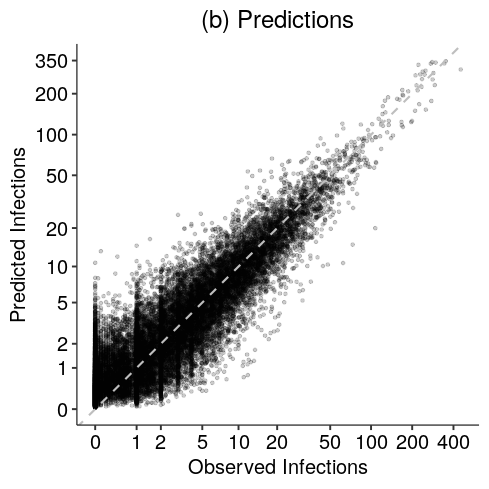}
		\includegraphics[width=0.65\linewidth]{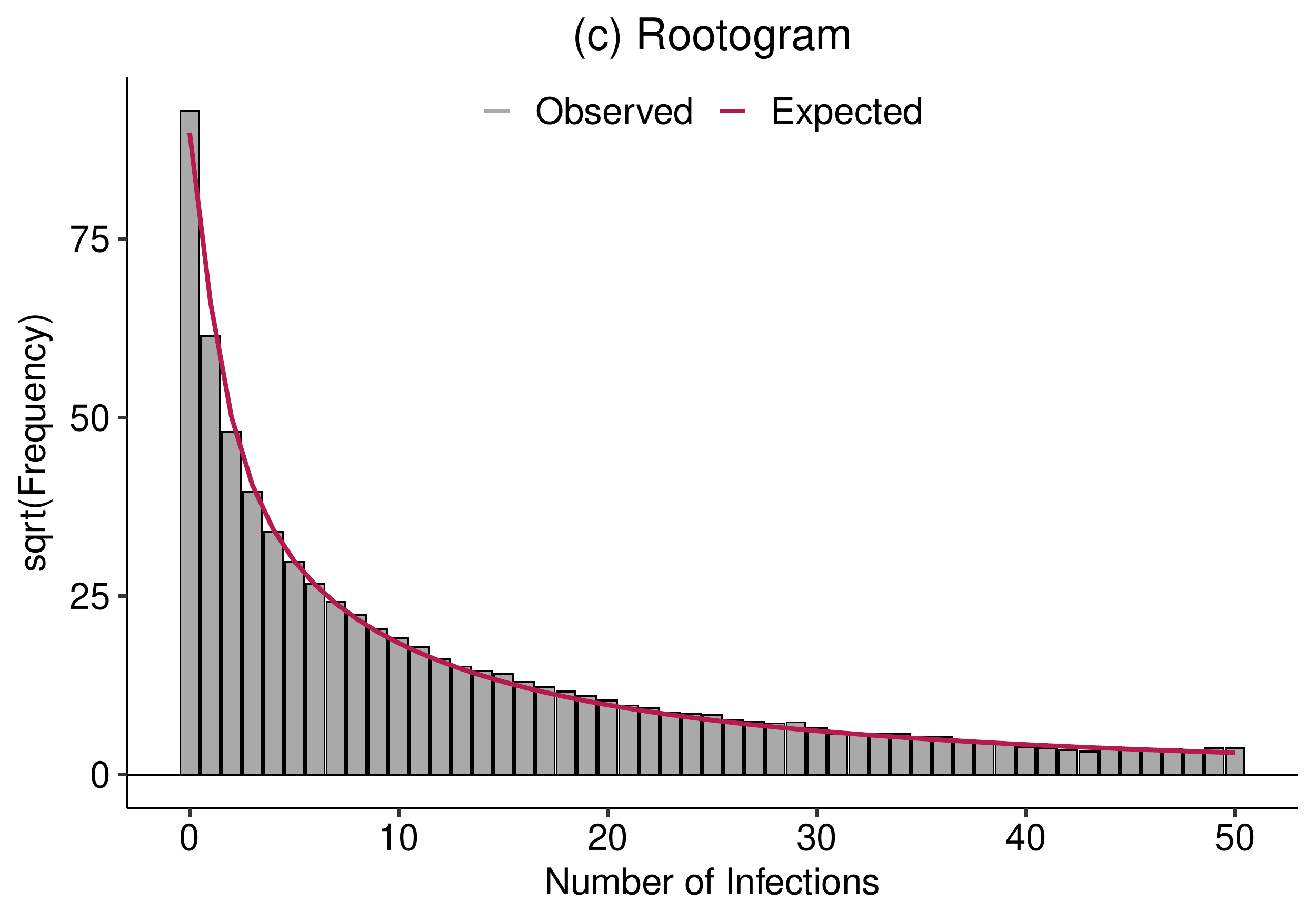}
		\caption{(a): QQ Plot of randomised quantile residuals, observations with a distance larger than 1 to the theoretically expected values are drawn in red. (b): Scatter plot of the observed and predicted infection count, for the x and y-axis, we used a $\log(\cdot + 1)$ scale. The dotted grey line is the best-case scenario of the prediction and has intercept 0 and slope 1. (c): Rootogram comparing the observed and expected counts. The grey barplot specifies the observed counts, while the red line gives the expected values under \eqref{eq:nb}.}
		\label{fig:gof}
	\end{figure}

	For further validation, we plot one draw of the randomised quantile residuals in Figure \ref{fig:gof} (a). \citet{Dunn1996} proposed this type of residual based on the result that evaluating the cumulative distribution function at all observed values of $y_{i,g,t}$ under the estimated parameters should yield uniformly distributed random variables. Transforming these uniform values by the quantile function of the standard normal gives the quantile residuals.  To obtain continuous residuals, the values are randomised since the negative binomial distribution in \eqref{eq:nb} has discrete support. On average, the empirical quantiles are close to the theoretical expectations and do not indicate problems regarding the statistical fit. At the right tail of the distribution, 38 (out of 24.060) observations exhibit higher deviations from the normal quantiles, which we coloured in red. The underlying counts are mainly credited to local outbreaks that could not be completely captured by the random effects, namely Coesfeld (Thuringia), Cuxhaven (Lower Saxony), Aichach-Friedberg (Bavaria), Guetersloh, and Warendorf (North Rhine-Westphalia). Additionally, we assess the predictions of the final model through plotting the predicted infections against the observed infections, Figure \ref{fig:gof} (b), and a rootogram proposed by \citet{Kleiber2016a}, Figure \ref{fig:gof} (b). Both visualisations confirm a strong fit of the presented model and proof that the model can sufficiently capture the observed counts of infected individuals. Due to the multiple imputation scheme specified in Section \ref{sec:imput} and \ref{sec:estimation}, we carry the model assessment out for each imputation separately and report the averaged results.

	\section{Conclusion}
	
	\label{sec:conclusion}
	In this writing, our contributions are twofold. Firstly, we used state-of-the-art regression models to quantify the importance of human mobility for understanding the spread of COVID-19 on a local level accounting for their temporal dynamic, latent effects and other covariates. Concerning the relative importance, the Gini index of meeting probability attribution proved to be a primary driver of the infection rates. Secondly, we used methods from multivariate statistics to derive friendship coordinates for the federal districts in Germany. Consecutively, we coupled the result with a regression model via isotropic splines and, thereby, revealed a  perpetual clustering of communities in former East- and West Germany, that remains existent for COVID-19 infections because the social geographic system proves to be an essential regressor in our application. Moreover, our findings enable an evaluation of the district-wise policies undertaken between March and June 2020. The results corroborate the usefulness of interventions limiting trans-district movements and concentrating meeting patterns.  Especially during the last weeks of this study, local lockdowns could mitigate further national outbreaks. 
	
	Still, we need to address some limitations of our work, which require additional investigation. The data sources for the infection data include all individuals in Germany that tested positive on COVID-19. During the peak phase in March, these tests were mainly carried out with patients who showed symptoms or had contact with an infected individual. Due to an unknown dark figure of infected persons missing in the public records \citep{Lavezzo2020a},  the observed data are a proxy for the current epidemiological situation. To control for this possible bias, further research on the prevalence of COVID-19 in Germany and the representability of the official statistics of the real infection occurrence akin to the REACT Study in England \citep{Riley2021} would be necessary. 
	
	Even with these caveats, the combination of infection, mobility, and connectivity data can serve for a fruitful application of other methods as well. Contrasting our approach, one may tackle the regression task in Section \ref{sec:model} by incorporating the spatial dependencies directly in the correlation structure, as is done in the literature on spatial econometric models \citep{LeSage2009}. We could also employ novel clustering algorithms that naturally exploit different proximity dimensions,  such as the geographical and social space, to identify similar districts while taking into account spatial dependencies \citep{DUrso2019,DUrso2020}. Further, the research questions posed in this article would greatly benefit from an examination through the lens of analytical sociology \citep{hedstroem2017}. Nevertheless, this type of analysis usually necessitates individual-level data, which are not readily available. Therefore, we can only verify some of the theoretical results of \citet{Block} on the macro scale, which does not necessarily translate to the micro scale \citep{stadtfeld2018_1}. Therefore, additional empirical work on the implications of individual behaviour on the spread of COVID-19 is still needed. Nevertheless, our work can give valuable pointers in that regard contingent on the assumption that the corresponding district average adequately represents the mobility patterns of an individual.

	\section{Data and code availability}
	
	Facebook collected all mobility data; however, we cannot share the raw data due to a data agreement. Still, we are allowed to provide all data aggregated onto the level of federal districts. To guarantee the replicability of our results, we make the complete code to obtain the results from this article available online. We also supply a visualisation of the entire pipeline of our analysis in the Supplementary Material for transparency.   
	
	\section*{Acknowledgement}
	We thank the anonymous reviewers for their careful reading and constructive comments. The project was supported by the European Cooperation in Science and Technology [COST Action CA15109 (COSTNET)]. This work has been funded by the German Federal Ministry of Education and Research (BMBF) under Grant No. 01IS18036A. The authors of this work take full responsibility for its content.

	\bibliographystyle{chicago}
	
	\bibliography{library_alt}

	\appendix

	\section{Multidimensional Scaling and Procrustes Analysis}
	\label{sec:mds}
	In order to determine the information given in the pairwise social connectedness indices  $x_{soc} = (x_{ij,soc})_{i,j = 1, ..., n}$ for explaining the spread of COVID-19 in Germany, we use techniques from multivariate statistics  \citep{Cox2000}. Thereby, we can derive a low-dimensional representation of the network on the actor level and guarantee interpretable as well as transparent results. More specifically, we apply \textsl{metric multidimensional scaling} (MDS) to represent dissimilarity matrices in a lower-dimensional geometric space that preserves the dissimilarities through euclidean distances \citep{Borg2013}. To illustrate the application of this algorithm, one can think of MDS as a technique to reverse-engineer geographic coordinates that are unique up to scale and rotation from distances between cities \citep{Young1938a}.
	
	At first, we transform the similarities expressed by the counts of friendship ties between the districts $x_{soc}$ to dissimilarities. In our application, the measure of dissimilarity is given by $d_{soc} = (\frac{1}{x_{ij,soc}})_{i \neq  j = 1, ..., n}$ and   $d_{ii,soc} = 0$. While this dissimilarity matrix is symmetric and nonnegative, there is no general guarantee that the entries of $d_{soc}$ are euclidean. Therefore, we add the constant $c$ to the off-diagonal elements to ensure that the  distances between the found coordinates are euclidean \citep{Cai1983, Mardia1978}.
	In order to estimate these $p$-dimensional coordinates $x_{i, soc} = (x_{i,1}, ..., x_{i,p}) ~\forall~ i = 1, ..., n$ from the dissimilarity matrix $d_{soc}$, the objective is to minimise the squared error between the pairwise entries of $d_{soc}$ and the euclidean distances calculated with the respective coordinates: 
	\begin{align}
	x_{soc} = (x_{1, soc}^\top, \ldots, x_{n, soc}^\top) = \underset{\tilde x \in \mathbb{R}^{p \times n}}{\text{argmin}} \Big( \sum_{i \neq j} (d_{ij,soc} + c- \lVert \tilde x_{i} -\tilde x_{j}\rVert^2) \Big)^{1/2},\label{eq:mds}
	\end{align} in our case we set $p = 2$. See \citet{Cox2000,Borg2013} for methods to find $x$ such that \eqref{eq:mds} holds, which are implemented in the R-package $\mathtt{stats}$ \citep{R}. 
	
	Since arbitrary transformations, rotations and reflections of any coordinates that optimise \eqref{eq:mds}, represented by $x_{soc} = \left( x_{1,soc}, ..., x_{n,soc} \right)$, are equally valid, we further process the solution to guarantee uniqueness and an intuitive understanding of the result. To achieve this goal, we use \textsl{Procrustes Analysis} \citep{Chen2008} to find an optimal solution $x_{soc}$ to \eqref{eq:mds} that is also most similar to the geographic coordinates $x_{coord} = \left( x_{1,coord}, ..., x_{n,coord} \right)$ given in Figure \ref{fig:mds}. As a measure of similarity between the matrices $x_{soc}$ and $x_{coord}$, commonly $R^2 = \sum_{i = 1}^n \left(x_{i,soc} -  x_{i,coord} \right)^\top \left(x_{i,soc} -  x_{i,coord} \right)$ is used. Further, we can parametrise the desired class of functions that transform an according to \eqref{eq:mds} optimal solution $x_{soc,i}$ to $\tilde{x}_{soc,i}$ by: 
	\begin{align}
	\tilde{x}_{soc,i} = \rho \mathcal{A}^\top x_{soc,i} + b,
	\end{align}
	where $\rho$ is scalar determining the dilation, $\mathcal{A}$ an orthogonal matrix defining the rotation and reflection, and $b$ a two-dimensional vector for a possible translation. From an optimisation point of view, we now have to find $\rho, \mathcal{A},$ and $b$ such that the resulting $R^2$ is minimised, which we can do in closed form (see \citealp{Cox2000}). This type of transformation is implemented in the R-package $\mathtt{vegan}$ \citep{vegan} and does not change the estimates or inference because we apply isotropic smooth terms. 
	
	%


	\section{Estimation of $\theta$ given $c$ and complete data}
	
	\label{sec:ann_estim}
	
	From \eqref{eq:nb} we construct a likelihood for each district and age/gender group tuple. Combining these separate contributions under independence leads to a joint logarithmic likelihood given by: 
	\begin{align}
	\ell(\theta,c) \propto \sum_{i = 1}^n \sum_{g \in \mathcal{G}}\sum_{t = 1}^T \log\left( \frac{\Gamma(\phi+y_{i,g,t})}{y_{i,g,t}!\Gamma(y_{i,g,t})}\right) + \phi \log\left(\frac{\phi}{\phi + \mu_{i,g,t}}\right)+ y_{i,g,t} \log\left(\frac{ \mu_{i,g,t}}{\phi + \mu_{i,g,t}}\right), 
	\label{eq:lh}
	\end{align}
	note that $\phi$ is the dispersion parameter of the negative binomial distribution and that the likelihood of the imputation model from \eqref{eq:negbin} in Section \ref{sec:imput} has the same form with $\phi^{-1} = \sigma_l$.   
	Suppose we plug $\mu_{i,g,t}$ as defined in \eqref{eq:end-epi} into \eqref{eq:lh} and  fix the value of $c$. In that case, we observe that the result is a function of $\theta$ and resembles the likelihood of a generalised additive model with negative binomial distributed target variables and denote the likelihood by $\ell(\theta | c)$ \citep{Ruppert2003a}.  To obtain a smooth fit of $\theta$, we extend this function by an additive penalisation component: 
	\begin{align}
	\ell_p(\theta | c) = \ell(\theta | c) - \tau^\top S, \label{eq:lh_pen} 
	\end{align}
	where $\tau = (\tau_{a},  \tau_{b}, \tau_{coord},\tau_{soc})$ are smoothing parameters weighting the term-specific penalties $S = ( S_a,  S_b, S_{coord},  S_{soc})$.
	The choice of these penalties differs between the random effects and bivariate spacial effects. For the random effects, we follow \citet{Ruppert2003a} and define $S_a$ and $S_b$ through ridge penalties, hence, for instance, $S_a =  \sum_{i = 1}^2 a_i^2$.   In the case of the isotropic semiparametric terms,  we chose the penalty terms in accordance with \citet{Duchon1977}. Here,   
	$S_{coord}$ penalises the roughness of the bivariate function $f_{coord}(x_{i,coord}) = f_{coord}(x_{i,coord, 1}, x_{i,coord, 2})$, where $x_{i,coord, p}$ denotes the $p$th dimension of $x_{i,coord} ~ \forall~ p \in \lbrace 1,2 \rbrace$, in our application the longitude and latitude of district $i$. Given this notation, we can state the functional form of the penalty term:
	\begin{align*}
	S_{coord} = \int &\frac{\partial^2}{\partial^2 x_{coord, 1}} f_{coord}(x_{coord})^2 + 2 \frac{\partial^2}{\partial x_{coord, 1}\partial x_{coord, 2}}f_{coord}(x_{coord})^2 + \\&\frac{\partial^2}{\partial^2 x_{coord, 2}} f_{coord}(x_{coord})^2  d x_{coord, 1}  d x_{coord, 2}.
	\end{align*} 
	Besides we ensure identifiability of all smooth effects by incorporating a sum-to-zero constraint per term, which translates to $\sum_{i = 1}^n f_{coord}(x_{i,coord}) = 0$ for $f_{coord}(\cdot)$ \citep{wood2017}. 
	

	To maximise \eqref{eq:lh_pen} in terms of $\theta$ and $\tau$, we follow the nested optimisation approach of \citet{Wood2011b}. Hence, we find $\hat{\tau}$ in an outer iteration and $\hat{\theta}$ consecutively in an inner iteration. Generally, the validity of this procedure rests on the finding that $\hat\theta$ is the posterior mode of $\theta|y$ under the assumption that $\theta$ follows a zero-mean normal prior with improper variance \citep{Kimeldorf1970}. Viewing $\theta$ as random coefficients enables us to estimate all smoothing parameters $\tau$ via restricted maximum likelihood estimation. More specifically, we set up $f(y,\theta| c)$ given $\ell(\theta| c)$ and $f(\theta)$. Through integrating $\theta$ out of $f(y,\theta| c)$ by deploying a Laplace approximation we obtain an approximate REML criterion, which is a function of $\tau$ and $\phi$, the dispersion parameter from \eqref{eq:lh}. Maximising the derived function in terms of these parameters gives $\hat{\tau}$ and $\hat{\phi}$ (see \citealp{Wood2011b} for additional details).  Given the tuning parameters, we consecutively find $\hat\theta$ through standard penalised iterative re-weighted least squares estimates (PIRLS, \citealp{wood2017}) in the inner iteration. We repeat this iterative scheme until convergence to obtain $\hat\theta$ and $\hat{\tau}$ given a fixed value of $c$. A scalable implementation of this routine that we used is available in the software package $\mathtt{mgcv}$ \citep{wood2017}.
     
\end{document}


\def\spacingset#1{\renewcommand{\baselinestretch}%
		{#1}\small\normalsize} \spacingset{1}

	
	\if1\blind
	{
		\title{\textbf{Supplementary Material \\ On the Interplay of Regional Mobility, Social Connectedness, and the Spread of COVID-19 in Germany}}
		\author{Cornelius Fritz and Göran Kauerman\hspace{.2cm}\\
			Department of Statistics, LMU Munich}
		\maketitle
		    \tableofcontents

	} \fi

	\if0\blind
	{
		\bigskip
		\bigskip
		\bigskip
		\begin{center}
			{\LARGE\bf Dynamic Networks}
		\end{center}
		\medskip
	} \fi
	
	\bigskip
      
   \newpage 
        \FloatBarrier
   
    \newpage
         \FloatBarrier

     \section{Analysis of Representativeness of Facebook Data}
    
    \begin{figure}[ht!]
    	\centering
    	\includegraphics[width=0.6\linewidth, page =1]{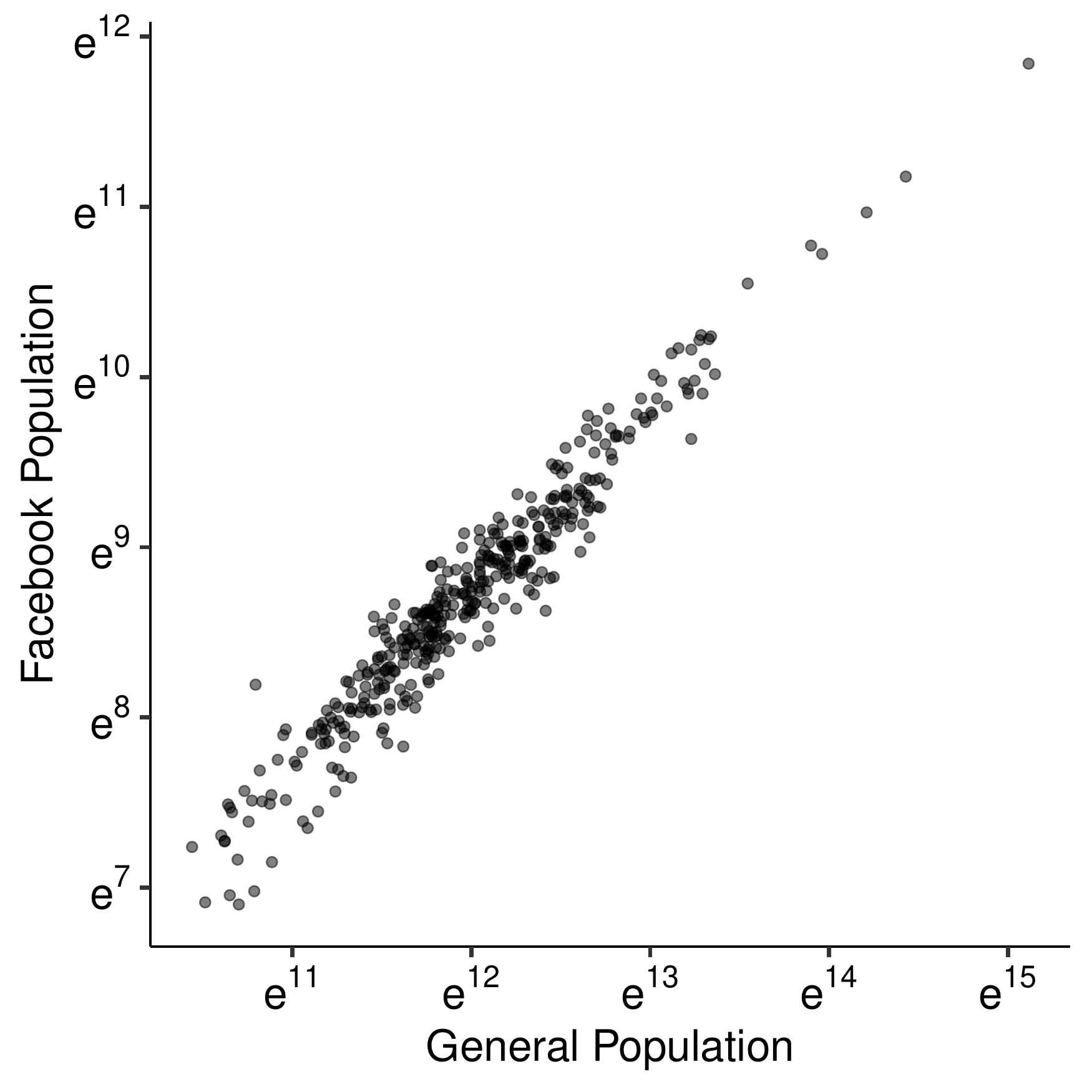}
    	\caption{Scatter-plot of the Facebook population that opted in the geolocation features and the general population for each district.}
    	\label{fig:pop_fb_vs_gen}
    \end{figure}

    To guarantee robust estimates, we investigate to what extent the mobility data provided by Facebook in the context of the  \textsl{Data for Good} program \citep{Maas,Iyer} are representative of the general German population and in line with other mobility data. We carry out the respective assessment in three parts: 
    \begin{enumerate}
    	\item Compare the spatial distribution of Facebook users that opted in geolocation features with the general population for each district.
    	\item  Contrast the age structure of Facebook users with the demographic pyramid in Germany. 
    	\item Check if mobility data from other providers, i.e., Apple and Google, measure similar information as the Facebook data. 
    \end{enumerate}
    
    \textbf{Spatial Distribution}:  Besides mobility data, the datasets made available by Facebook also include the daily counts of Facebook users who enabled geolocation features in each federal district. For this assessment, we only look at the average of users per district over the study period from the 3rd of March to the 16th of June 2020 and calculate the same quantity for the general population provided by the  \href{https://www.destatis.de/EN/Themes/Society-Environment/Population/Current-Population/_node.html}{German Federal Statistical Office}.
    Figure \ref{fig:pop_fb_vs_gen} is a scatter-plot of 401 points representing the general and Facebook population measured for each federal district. We can conclude from this plot that the spatial distribution of Facebook users is positively correlated with official statistics, i.e., the estimated Pearson correlation is 0.98. 
    
    \begin{figure}[t!]
    	\centering
    	\includegraphics[width=0.6\linewidth, page =1]{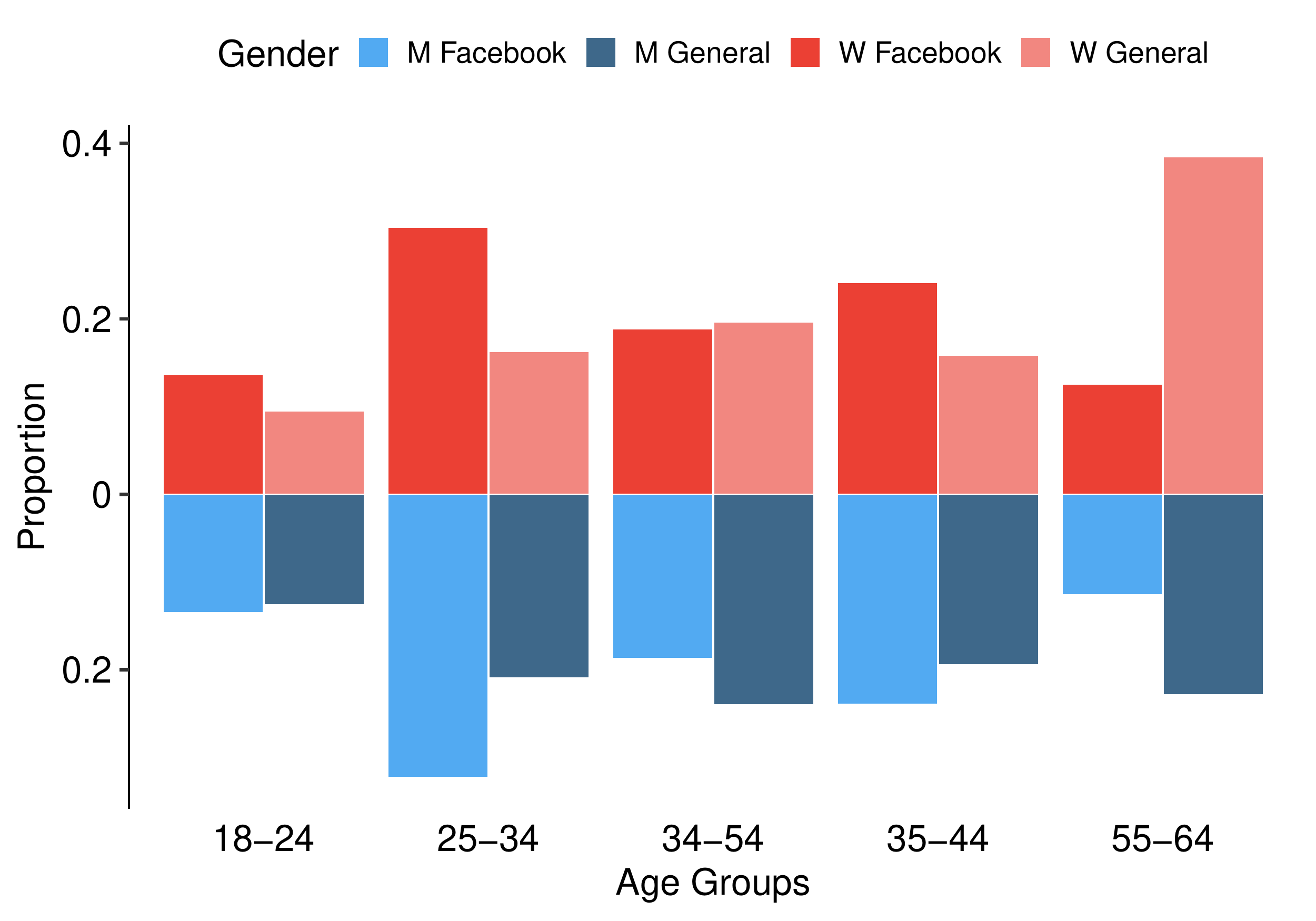}
    	\caption{Comparison of the age- and gender-structure within Facebook and in the general population of Germany. }
    	\label{fig:age_fb_vs_gen}
    \end{figure}
    
    \textbf{Age Representation}: In addition to people's spatial distribution, we want to compare the age structure of general Facebook users and the general German population. The Facebook users' age structure that turned on the geolocation features is unknown due to differential privacy. As a possible proxy for this information, we compile information about the age structure of the entire Facebook population in Germany from \href{https://www.facebook.com/ads/audience-insights}{Facebook Audience Insights}. Unfortunately, Facebook's age-groups are not following the age groups given by the RKI. 
    As can be seen in Figure \ref{fig:age_fb_vs_gen}, the Facebook population is younger than the general population. While we can conclude that the age groups 18-24, 35-54 are adequately represented in the Facebook sample, there is a surplus of people in the 25-to-34-year-old cohort. Simultaneously, the age group of the oldest individuals is in relative terms less populated than the general population. Still, this finding does not necessarily indicate a bias in the mobility data that we use in the principal analysis. Due to the standardisation of the mobility-related covariates given in Formula (1) of the main paper, the covariates are robust to a nation-wide under- and over-representation of older and younger individuals. In that case, the measurement bias would only be a constant, and hence the standardised covariates would be the same. We cannot quantify with the available data whether the percentage compositions of those age structures vary within each district. However, the promising results of the spatial representativeness make it reasonable to assume that this is not the case. 
    
    \begin{figure}[t!]
    	\centering
    	\includegraphics[width=0.7\linewidth, page =1]{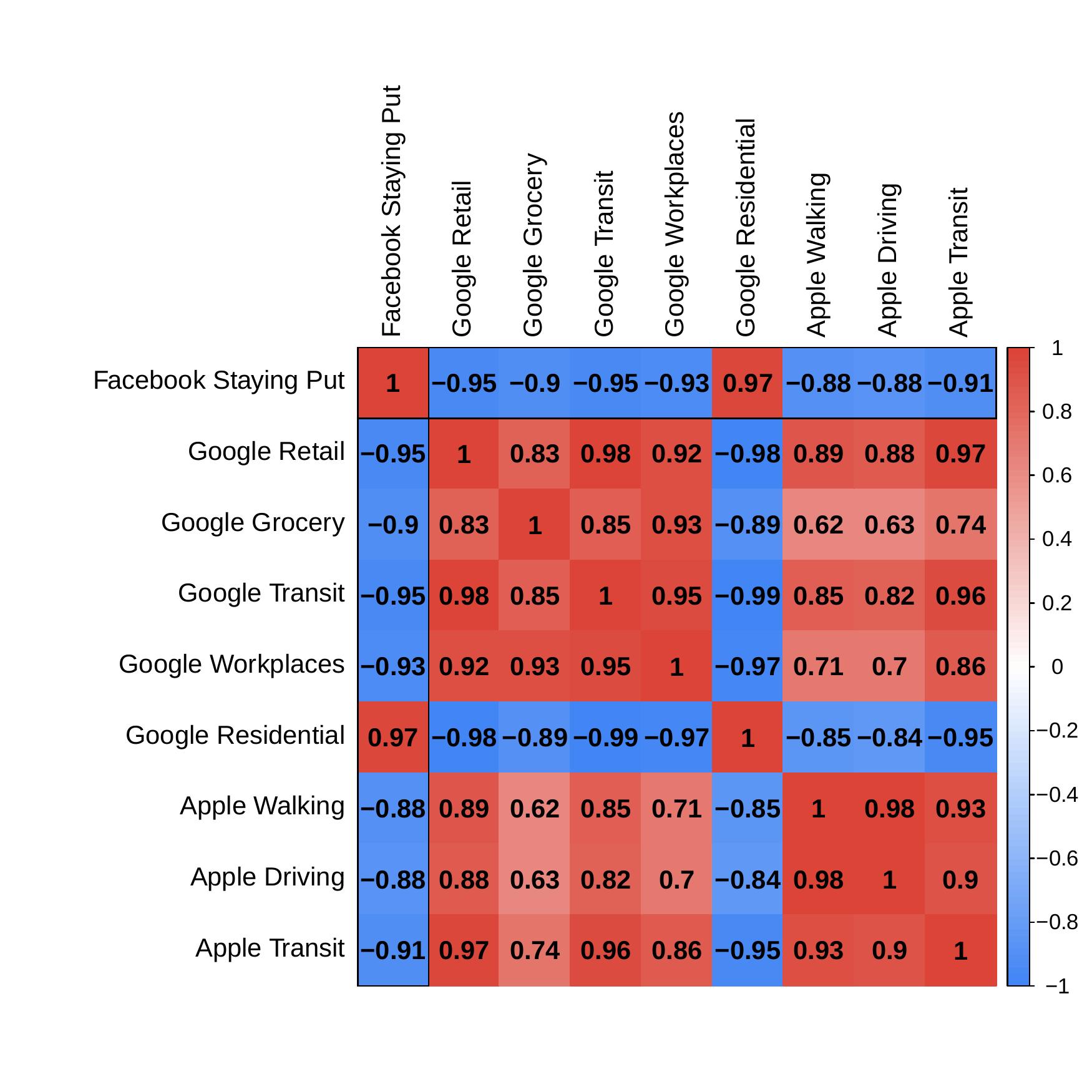}
    	\caption{Correlation matrix of the weekly national mobility indicators from Facebook, Google, and Apple.}
    	\label{fig:corrmat}
    \end{figure}

    \textbf{Mobility Measurement from other Sources}: We now contrast the mobility data provided by two other major technology companies, namely Apple\footnote{\url{https://covid19.apple.com/mobility}} and Google\footnote{\url{https://www.google.com/covid19/mobility/}}, with the data made available by Facebook. We use the Google Community Mobility Reports, including relative mobility trends in retail or grocery stores, transit stations, and places of work. Similarly, Mobility Trends Reports from Apple use information on the relative requests for directions to walk, drive or use transit transportation in Apple Maps to measure the mobility trends. Since solely the Percentage of People Staying Put relates to absolute movements in the Facebook data, we investigate the Pearson correlation matrix of all given mobility indices in Figure \ref{fig:corrmat}. The high correlations in the first row and column are conclusive in that the Percentage of People Staying Put captures the information of all other variables reasonably well since the absolute correlation coefficients range from 0.88 to 0.97. 
    
    
    
    \section{Pipeline of the Analysis}

    \begin{figure}[!]
    	\centering
    	\includegraphics[clip,width=\linewidth]{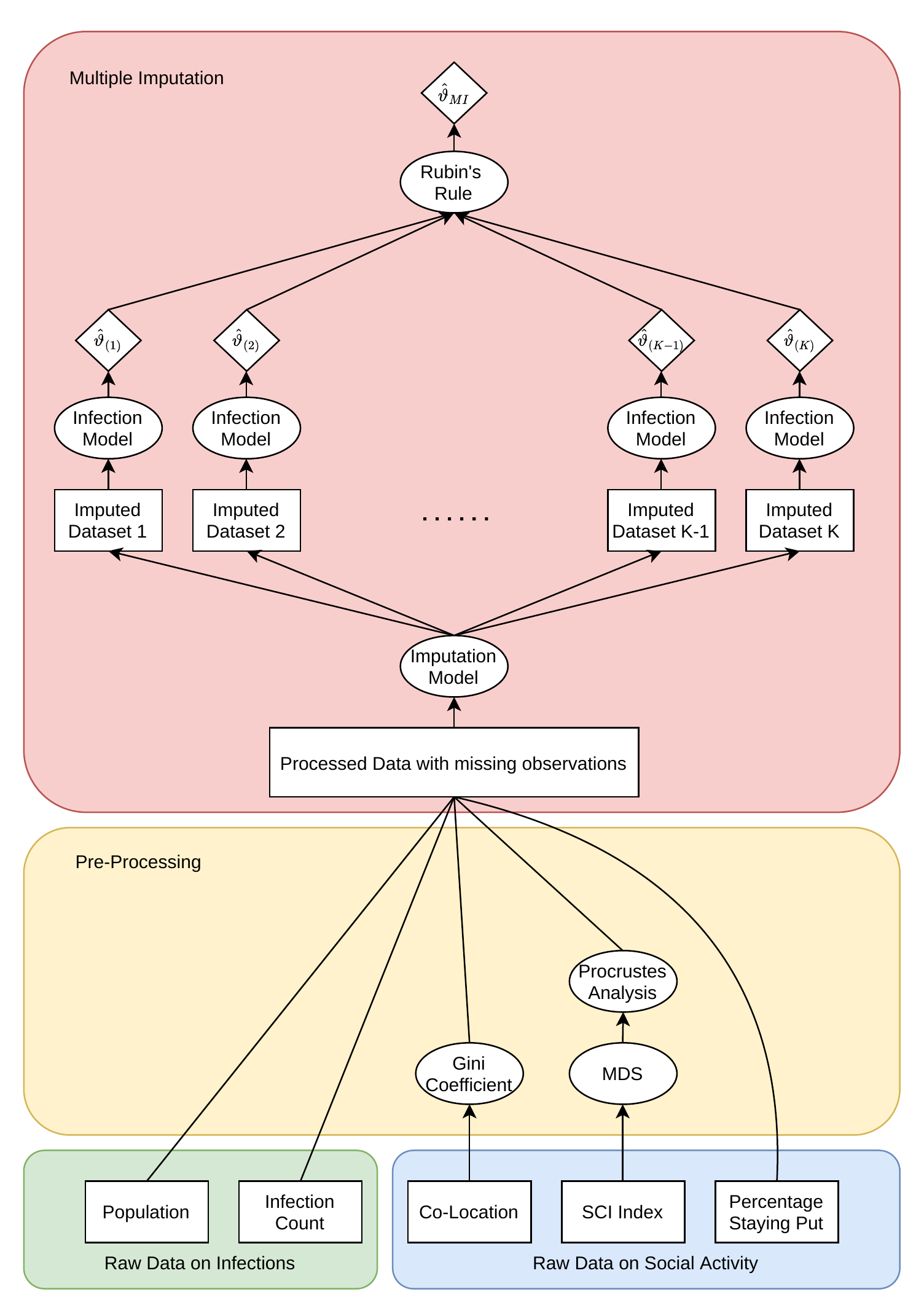}
    	\caption{Visualisation of the complete pipeline from pre-processing raw data to obtaining the final estimates. Rectangles relate to data, ovals to procedures, and tilted squares to estimators.}
    	\label{fig:vis}
    \end{figure}
    To make the performed analysis of the main article as transparent as possible, Figure \ref{fig:vis} depicts the complete pipeline, including all stages needed for this work. Information on the raw data and needed pre-processing steps are given in Section 2, Section 3 and Annex A. The imputation and infection models, together with Ruben's formula to pool estimates from different imputations, are specified in Section 3 (all referred contend relates to the main article).

    \section{Imputation of the Test Delay}

    The imputation procedure given in the main article relies on all observations to be missing at random (MAR,\citealp{Little2002}). To check this assumption, we first argue for stochasticity of the mechanism driving the binary indicator whether the target variable, i.e., the date of disease onset, is missing. In a second step, we perform a missing at random analysis to discover which covariates affect this mechanism. We consecutively employ the detected covariates in the imputation procedure introduced in Section 2.3 of the main article and report the full estimates. Finally, we conduct a sensitivity analysis to check whether our findings change if we only regard cases with observed disease onset.
    
    \subsection{Stochasticity of Missings} 
    
    \begin{figure}[!]
    	\centering
    	\includegraphics[trim={0.5cm 0.7cm 0.5cm 1cm},clip,width=0.7\linewidth, page =3]{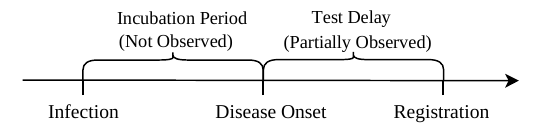}
    	\caption{Illustrative temporal path of a COVID-19 infection, collected surveillance data and resulting delays. }
    	\label{fig:observations}
    \end{figure}
    
    Each registered COVID-19 case can be characterised by three dates representing distinct stages of the disease, namely the time of infection, disease onset, and registration at local health authorities. This progression is illustrated in Figure \ref{fig:observations}. While we do not observe the disease onset date for some cases, we can still argue that all samples progress through the same three stages, although the second date is in some cases latent. Hence under the assumption that the incubation period from the infection date to the (observed or latent) disease onset date is stochastic, it is also appropriate to assume that the same holds for the test delay, defined as the time-span between disease onset and registration date. Given this argument, it is legitimate to impute the test delay through a stochastic model. We can then use this information to project the disease onset date back from the registration date for all cases given the observed data.  Besides, we carry out a sensitivity analysis in Section \ref{sec:sens} to check for structural differences in the dynamics of infection rates of cases with a disease onset and all cases. The estimates excluding all observations with missing disease onset date are in accordance with our findings of the main article.

    \subsection{Missing at Random Analysis}
    \label{sec:maranalysis}
    
    \begin{table}[!]
    	
    	\label{tbl:specs}
    	\center
    	\begin{tabular}{l|c c c c c c |l c} 
    		& Time Trend & Age & Gender &   Age:Gender & District & State & cAIC & Rank \\ \hline
    		Model 1 &  $\checkmark$    &      $\checkmark$      &      $\checkmark$     & $\checkmark$ &  $\checkmark$&  $\checkmark$&  121608.4 & 1  \\
    		Model 2 &  $\checkmark$    &      $\checkmark$      &      $\checkmark$     & $\checkmark$ &  $\checkmark$&  &   121610.7  & 2  \\
    		Model 3 &  $\checkmark$    &      $\checkmark$      &      $\checkmark$     & $\checkmark$ &  &  $\checkmark$&   142637.8& 5 \\
    		Model 4 &  $\checkmark$    &      $\checkmark$      &      $\checkmark$     &  &  $\checkmark$&  $\checkmark$& 126562  & 7\\
    		Model 5 &  $\checkmark$    &      $\checkmark$      &          & $\checkmark$ &  $\checkmark$&  $\checkmark$&  121632.1 & 6\\
    		Model 6 &  $\checkmark$    &           &      $\checkmark$     & $\checkmark$ &  $\checkmark$&  $\checkmark$& 121741.9 & 4  \\
    		Model 7 &      &      $\checkmark$      &      $\checkmark$     & $\checkmark$ &  $\checkmark$&  $\checkmark$&  121731.1 & 3 \\
    	\end{tabular} 
    	\caption{Different specifications of the missing at random process. The sign $\checkmark$ signals the inclusion of a specific covariate in the respective model. In the last two rows the corresponding corrected AIC (cAIC) values and their rankings are given.}
    	\label{tbl:mar}
    \end{table}

    Having established that the missing values are random, we need to identify the covariates driving the missing values' mechanism. To do that, we generate a binary indicator of whether the test delay was observed for each case $l$. In the consecutive step, we take this binary indicator to be the target variable of a logistic regression with seven different sets of covariates given in Table \ref{tbl:mar}, which we compare utilising the corrected AIC values \citep{wood2016}.  Further, we include all categorical effect by dummy-coding them. All effects are fixed beside the Gaussian random district-specific effects.  We parametrise the temporal trend by a penalised spline \citep{eilers1996}.  In the final row of Table \ref{tbl:mar}, the corresponding cAIC values indicate that Model 1 is the best specification of the missing at random process; hence we employ the corresponding set of covariates in our imputation model.

    \subsection{Results}
    
    \begin{table}[t!]
    	\begin{center}
    		\begin{tabular}{l | c c}
    			\multirow{3}{2cm}{Covariable} & \multicolumn{2}{c}{Parameter} \\
    			&  $\mu_{l}$ & $\sigma_{l}$\\
    			&  (Standard Error) &  (Standard Error) \\
    			\hline
    			Intercept &  1.852 & $-1.011$ \\
    			& $(< 0.01)$ & 0.022 \\
    			Male                    & $0.005$ &  $-0.069$ \\
    			& $(< 0.01)$ &   $(0.024)$   \\
    			A35-A59            & $0.017$& $0.101$  \\    			
    			& $(< 0.01)$     &$(0.02)$ \\
    			A35-A59:Male           & $0.038$& $-0.027 $  \\
    			& $(< 0.01)$     &$(0.030)$ \\
    			Weekend     &           $-0.034$   & $-0.049$  \\
    			& $(< 0.01)$   &   (0.017) \\
    		\end{tabular}
    		\caption{Estimates of linear effects from the imputation model.The reference group are female individuals aged between 15 and 35 living in Baden-Württemberg.}
    		\label{table:coefficients_gamlss}
    	\end{center}
    \end{table}

    Originating from the missing at random analysis of Section \ref{sec:maranalysis}, our imputation model incorporates a random district-specific effect, a temporal trend, and dummy covariates for the age, gender, and state cohort as well as an interaction between age and gender groups. In our proposed imputation procedure, we parametrise not only the mean of the test delay ($\mu$) but also its scale factor ($\sigma$). Therefore, we provide the full estimates of the corresponding model defined in Formula (2) of the main article separately for each coefficient. Besides, we decompose all reported effects into sociodemographic, state, smooth and random effects. 
    
    \textbf{Sociodemographic Effects}: The results of demographic terms are given in Table \ref{table:coefficients_gamlss} and should be interpreted regarding the mean and dispersion parameter of the period between disease onset date and reporting date, which we define as test delay.

\begin{figure}[t!]
    	\centering
    	\includegraphics[width=0.6\linewidth, page =1]{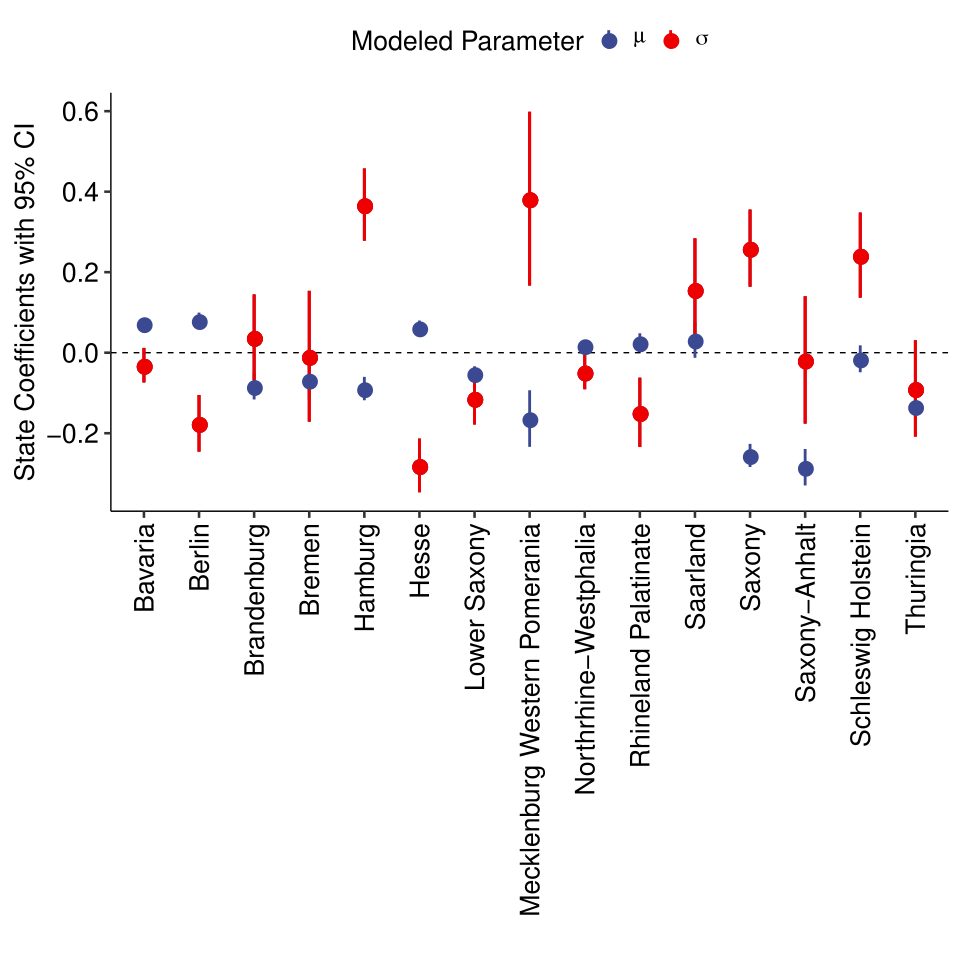}
    	\caption{Linear State Effects of the test delay model on $\mu$ and  $\sigma$. The reference category is Baden-Württemberg, hence the respective coefficient is fixed at zero. }
    	\label{fig:fixed_effects_imputation}
    \end{figure}

    \textbf{State Effects}: In addition to the demographic covariates discussed in the main article, we included a fixed effect for each state. For the results shown in Figure \ref{fig:fixed_effects_imputation}, we use Baden-Württemberg as the reference category. Hence, all estimates should be interpreted relative to Baden-Württemberg. The state effects on $\mu$ in Figure  \ref{fig:fixed_effects_imputation} (a) indicate that the average test delay in eastern states, e.g., Saxony-Anhalt and Saxony, is lower than in the reference category. The test delay volatility also varies significantly between the states, as shown in Figure \ref{fig:fixed_effects_imputation} (b), e.g., it is the highest in Mecklenburg Western Pomerania and Hamburg.     
    
    \textbf{Smooth Effects}:  We observe a negative temporal trend for reporting dates during the beginning of March. But once we see more reported cases in April, the average test delay lengthens. During this period, the $\sigma$ parameter is higher, leading to a higher variance of the respective observations. After May, the average delay decreases but increases again during the last week of this study. 
    
    \begin{figure}[t!]
    	\centering
    	\includegraphics[width=0.45\linewidth, page =1]{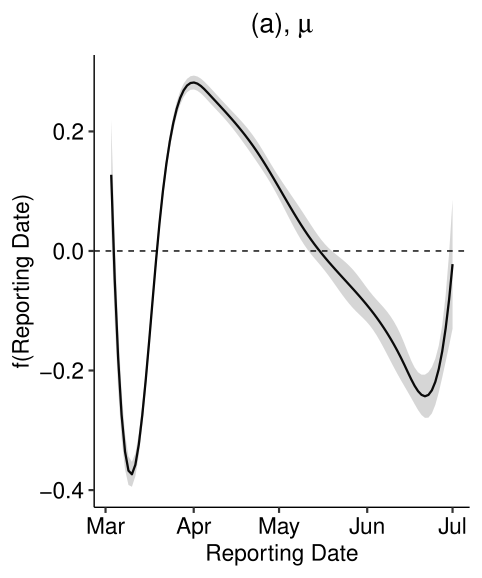}
    	\includegraphics[width=0.45\linewidth, page =1]{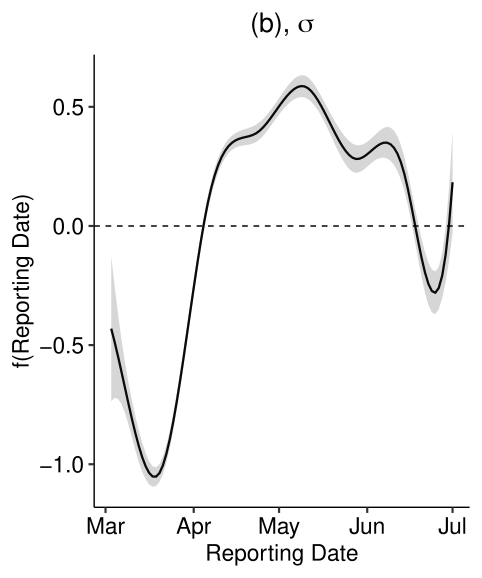}
    	\caption{Smooth temporal trends of the $\mu$ (a) and  $\sigma$ (b) parameter of the test delay model.}
    	\label{fig:estimates_onset}
    \end{figure}
    
\begin{figure}[t!]
    	\centering
    	\includegraphics[width=0.9\linewidth, page =1]{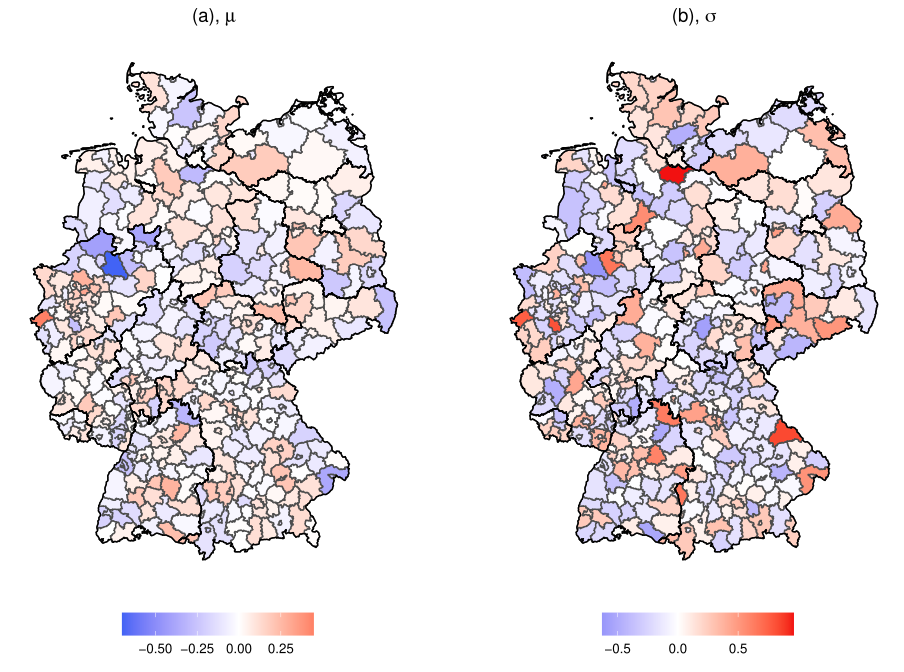}
    	\caption{Random district-specific effects regarding $\mu$ (a) and  $\sigma$ (b) parameter of the test delay model.}
    	\label{fig:random_effects_imputation}
    \end{figure}

    \textbf{Random Terms}: The district-specific intercept effects for $\mu$ are given in Figure \ref{fig:random_effects_imputation} (a), while Figure \ref{fig:random_effects_imputation} (b) shows them for $\sigma$.

    \newpage
    \subsection{Sensitivity Analysis of Disease Onset Imputation}
    
    \label{sec:sens}
\begin{figure}[t!]
    	\centering
    	\includegraphics[width=0.4\linewidth, page =1]{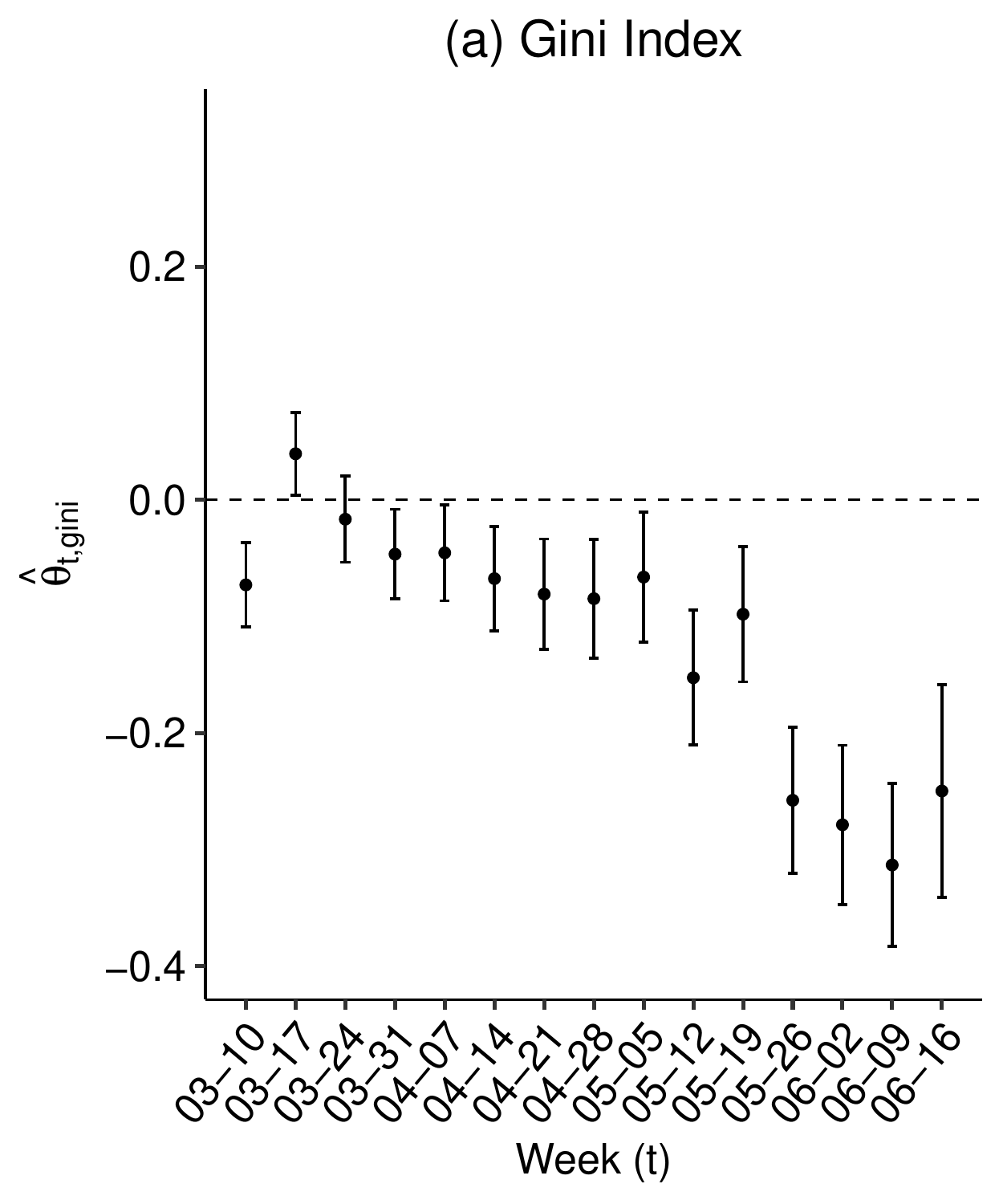}
    	\includegraphics[width=0.4\linewidth, page =1]{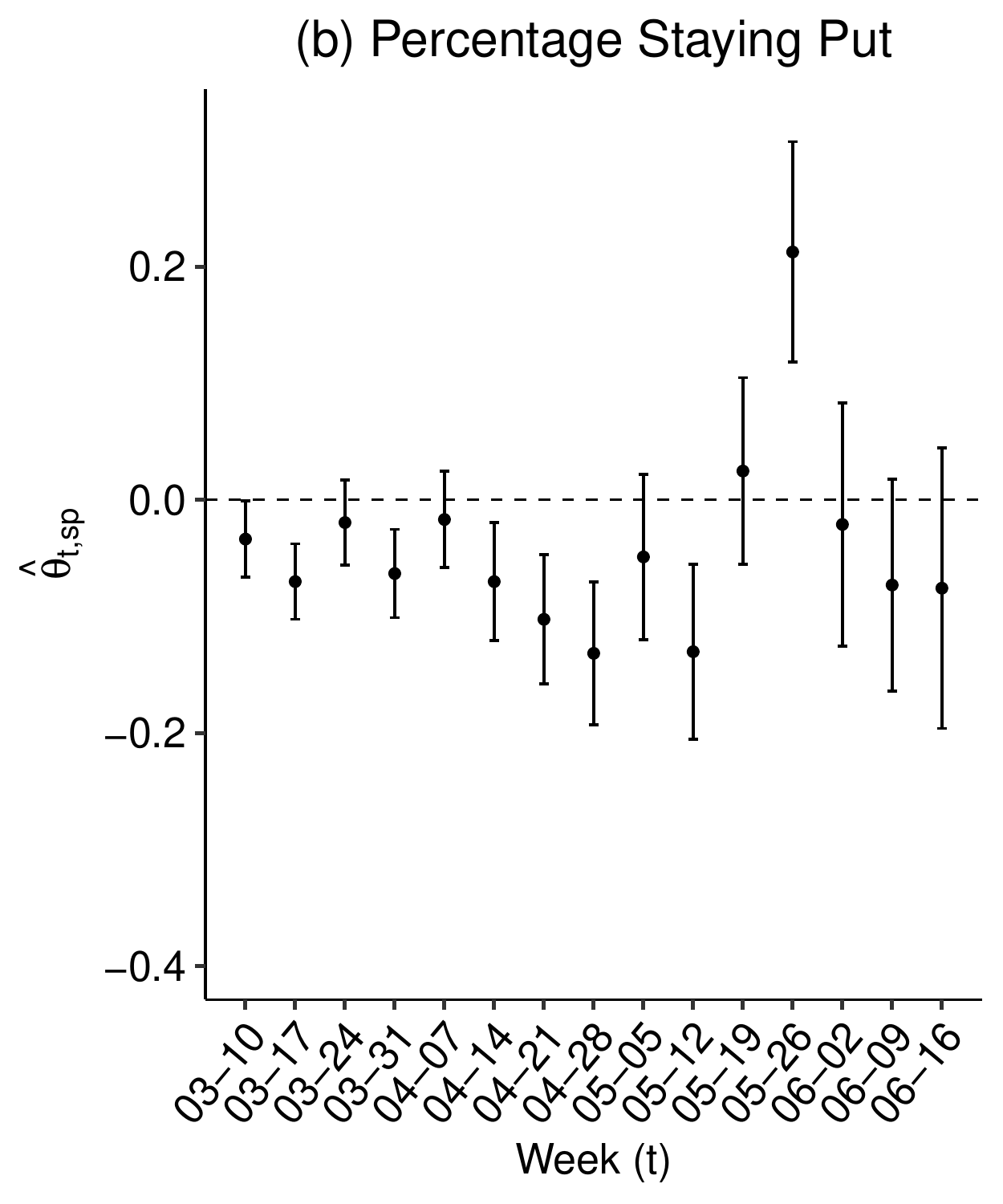}
    	\caption{(a): Time-varying effects of the Gini index $\hat{\theta}_{t,gini}$. (b): Time-varying effects of the Percentage of People Staying Put $\hat{\theta}_{t,sp}$. The $95\%$ confidence interval accompanies the estimates, and the shown dates (mm:dd)  on the x-axis are the first days of the corresponding weeks.  All cases with missing disease onset date are excluded.}
    	\label{fig:beg}
    \end{figure}

\begin{figure}[t!]
    	\centering
    	\includegraphics[width=0.75\linewidth, page =2]{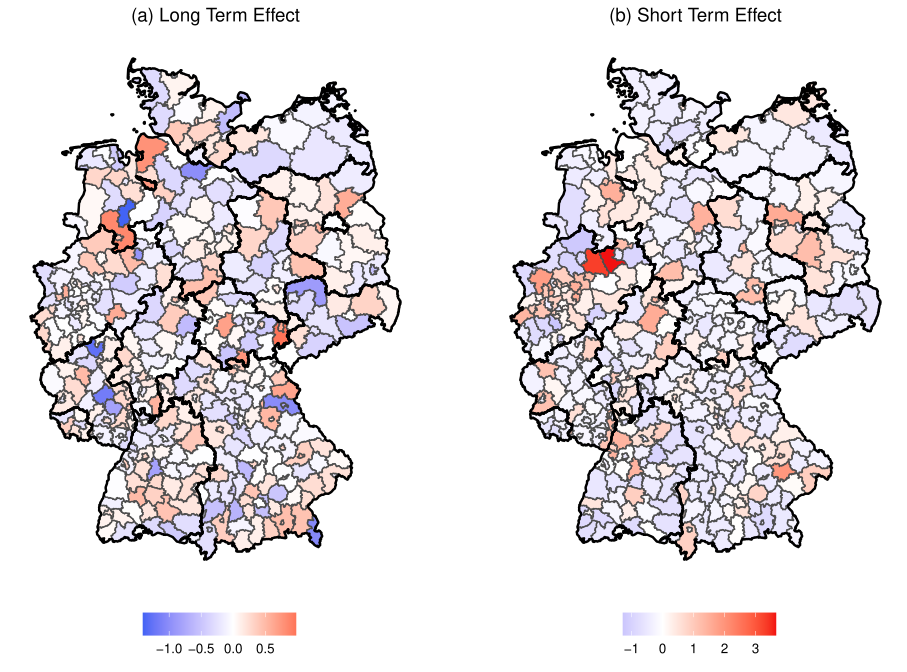}
    	\caption{(a): Maximum posterior modes of the long-term random effects $a_i$. (b) Maximum posterior modes of the short-term random effects $b_i$. The thick black lines represent borders between federal states, while the thinner grey borders separate federal districts. All cases with missing disease onset date are excluded.}
    	\label{fig:end}
    \end{figure}

    We carry out a sensitivity analysis of the imputation procedure introduced in Section 4 of the main manuscript to check whether our findings differ if we only regard cases with observed disease onset. In Figures \ref{fig:beg} to \ref{fig:end}, the model results are given if we only include the cases where the disease onset date is recorded in the surveillance data.  All findings of the principal analysis are robust to the imputation method used; hence we see no structural differences that arise due to including all cases rather than only the cases with an observed disease onset.  
    
    \begin{figure}[t!]
    	\centering
    	\includegraphics[width=0.45\linewidth, page =1]{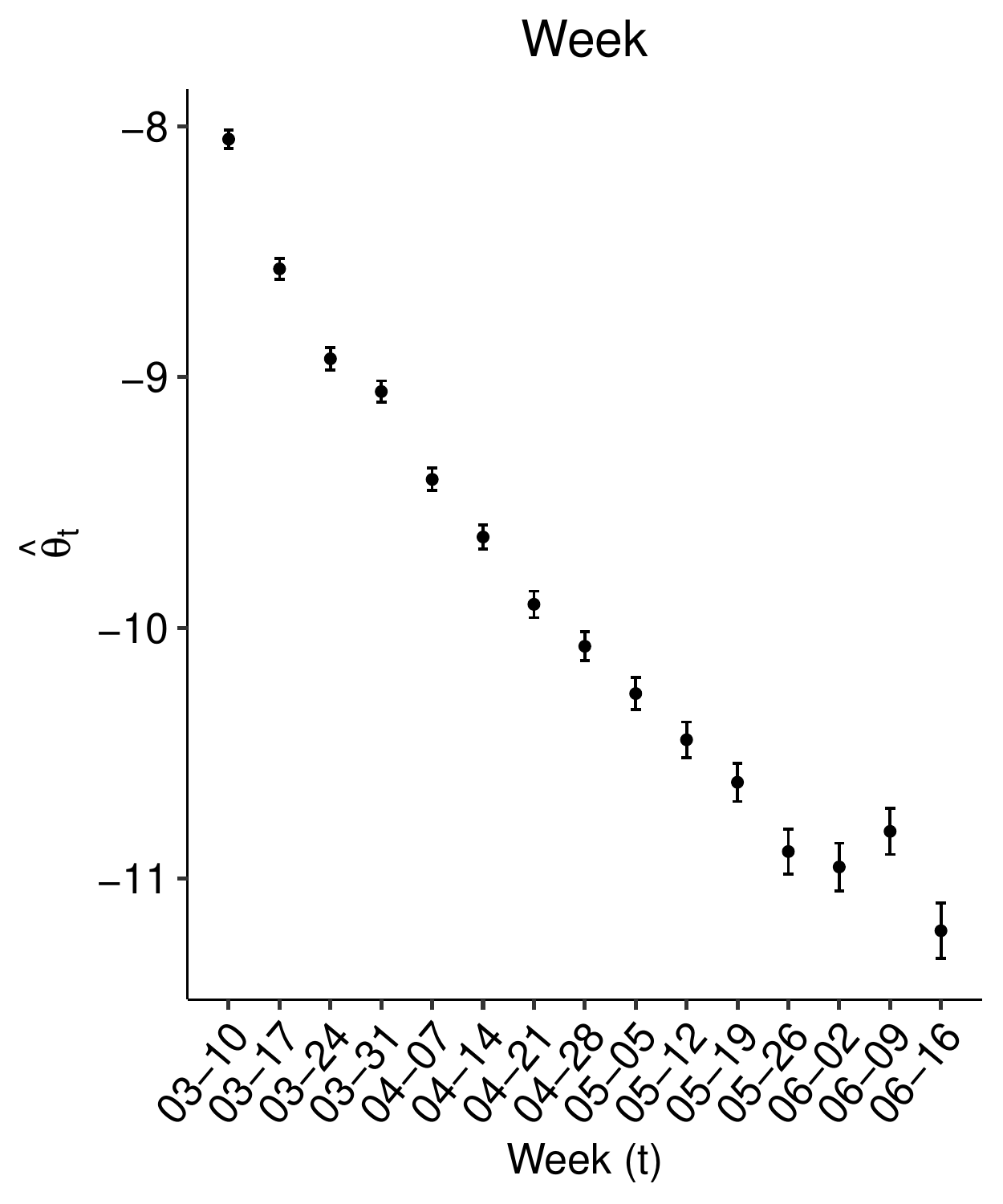}
    	\caption{Estimate of temporal effect $\theta_t$. The $95\%$ confidence interval accompanies the estimates, and the shown dates (mm:dd)  on the x-axis are the first days of the corresponding weeks. All cases with missing disease onset date are excluded.}
    	
    	\label{fig:base2}
    \end{figure}

    \begin{figure}[t!]
    	\centering
    	\includegraphics[width=0.75\linewidth]{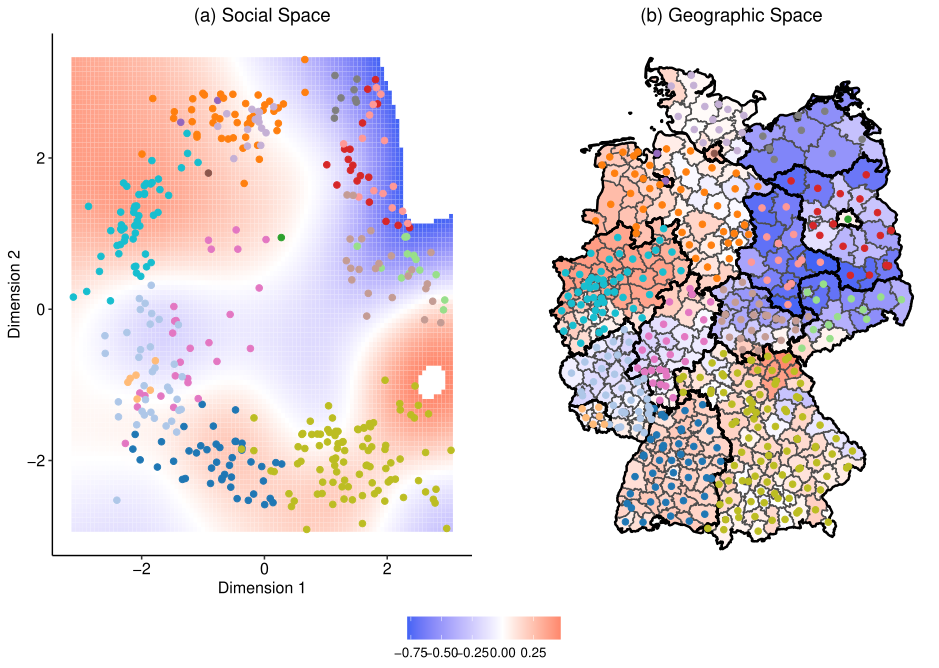}
    	\caption{(a): Coordinates of the districts in the friendship space with the smooth partial effect of $f_{soc}$ in the background. We only show the predictions in the range of observed values. (b):  Coordinates of the districts in the geometric space with the smooth partial effect of $f_{soc}$ again shown in the background for each district. The thick black lines represent borders between federal states, while the thinner grey borders separate federal districts. All cases with missing disease onset date are excluded.}
    \end{figure}
    
    \begin{figure}[t!]
    	\centering
    	\includegraphics[width=0.35\linewidth, page =1]{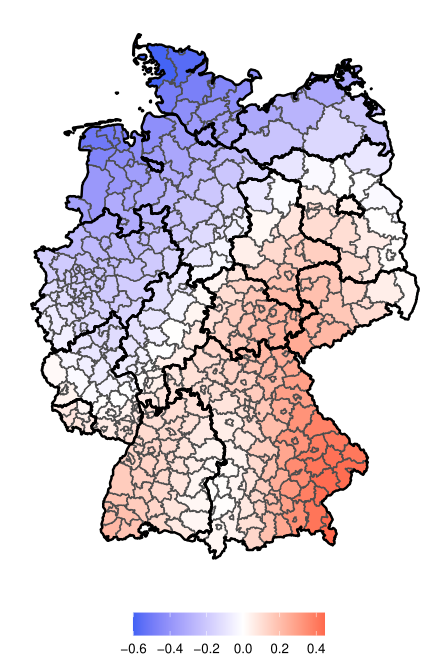}
    	\caption{Estimated smooth spatial effect $f_{coord}$. The thick black lines represent borders between federal states, while the thinner grey borders separate federal districts. All cases with missing disease onset date are excluded.}
    \end{figure}

    \section{Alternative Infection Model}
    
\begin{figure}[t!]
    	\centering
    	\includegraphics[width=0.4\linewidth]{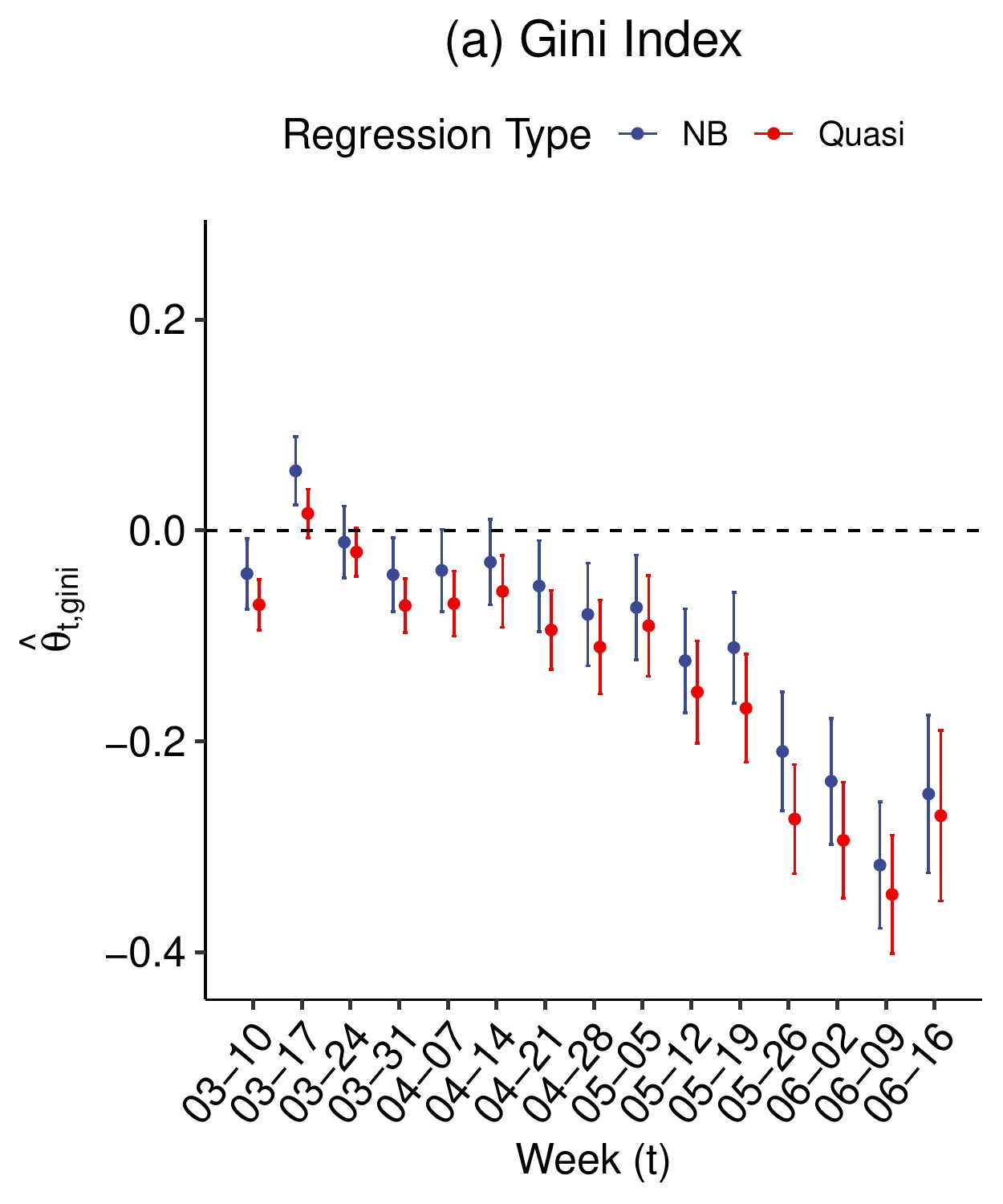}
    	\includegraphics[width=0.4\linewidth]{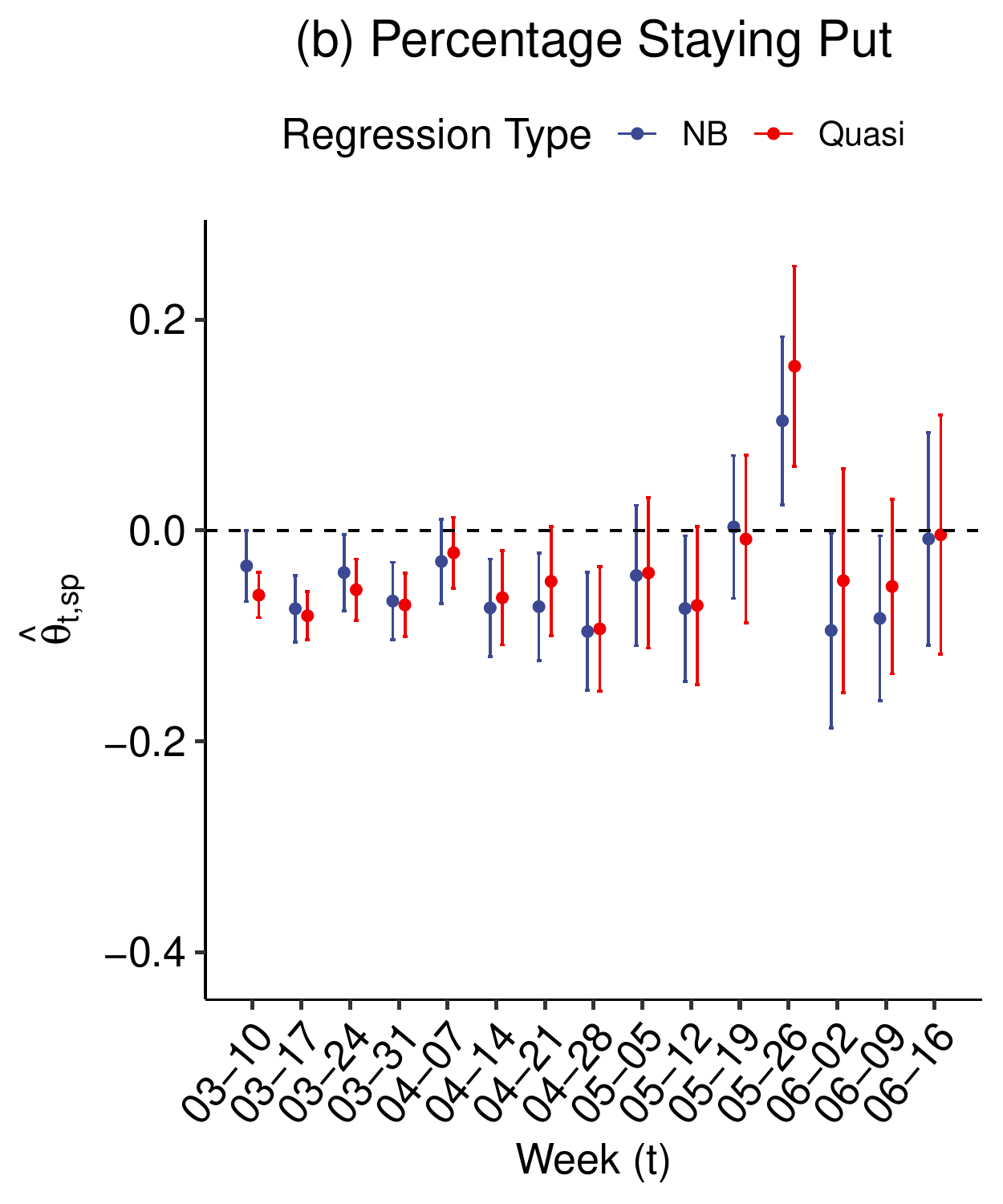}
    	\caption{(a): Time-varying effects of the Gini index $\hat{\theta}_{t,gini}$ in the Quasi-Likelihood and the Negative Binomial model. (b): Time-varying effects of the Percentage of People Staying Put $\hat{\theta}_{t,sp}$ in the Quasi-Likelihood and the Negative Binomial model. All estimates are accompanied by $95\%$ confidence interval and the colour indicates whether the coefficients relate to the original Negative binomial fit presented in the main analysis or the quasi likelihood approach.}
    	\label{fig:quasi1}
    \end{figure}

\begin{figure}[t!]
    	\centering
    	\includegraphics[width=0.4\linewidth]{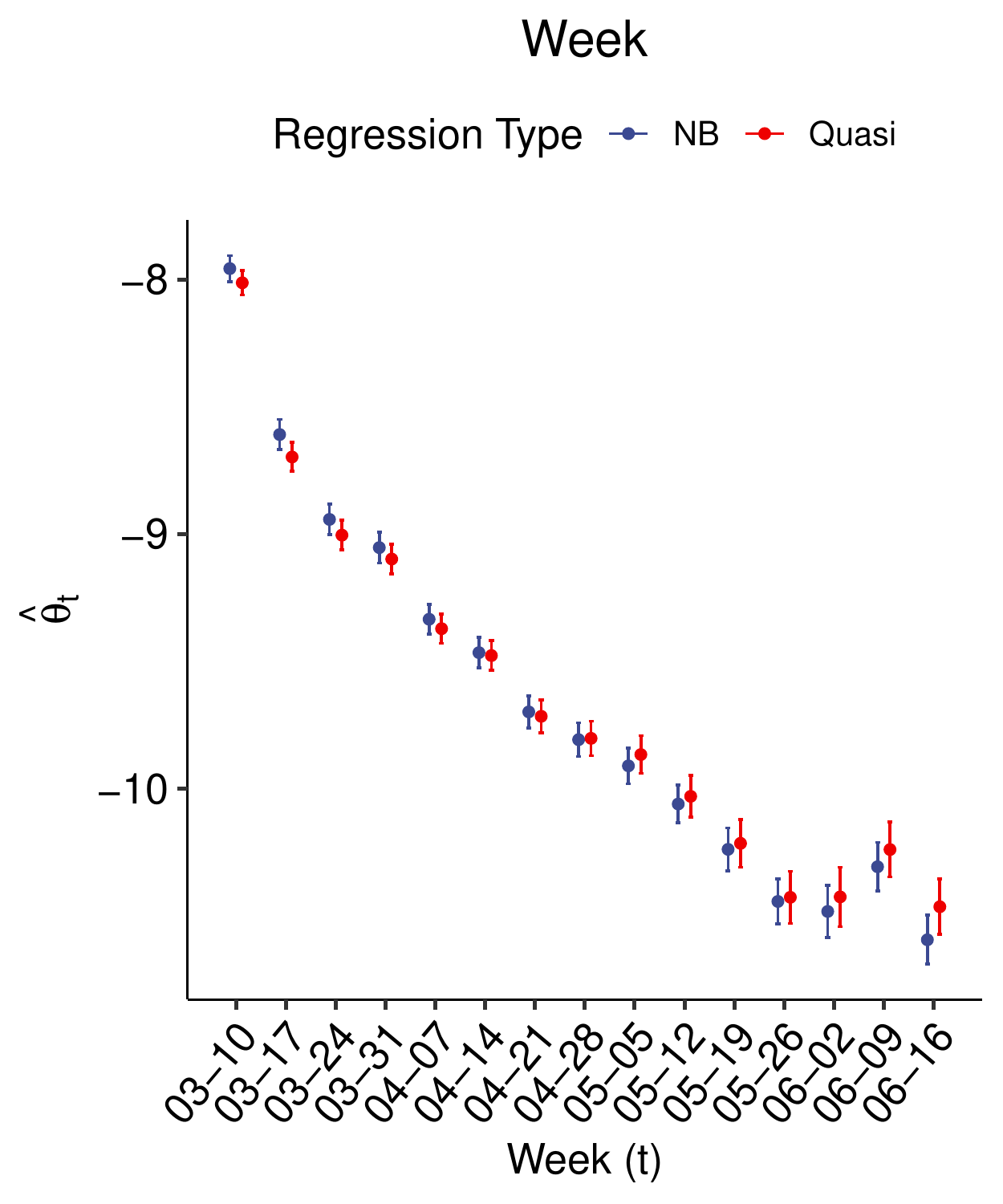}
    	\caption{Time-varying baseline intensity $\theta_t$ in the Quasi-Likelihood and the Negative Binomial model. The estimates are accompanied by the $95\%$ confidence interval and the colour indicates whether the coefficients relate to the original Negative binomial fit presented in the main analysis or the quasi likelihood approach. }
    	\label{fig:quasi2}
    \end{figure}

\begin{figure}[t!]
    	\centering
    	\includegraphics[width=0.75\linewidth]{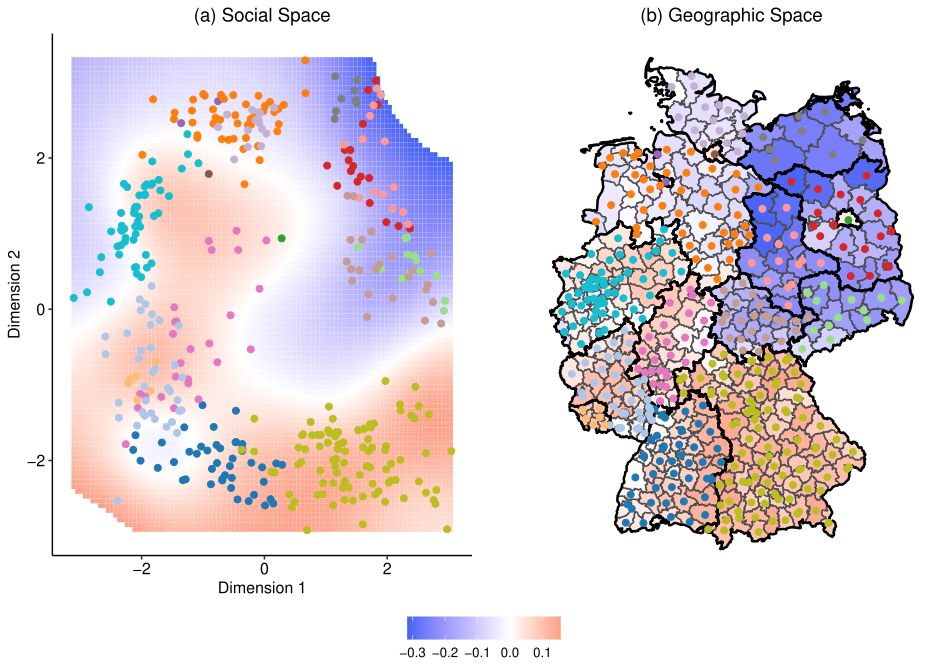}
    	\caption{(a): Coordinates of the districts in the friendship space with the smooth partial effect of $f_{soc}$ in the background. We only show the predictions in the range of observed values. (b):  Coordinates of the districts in the geometric space with the smooth partial effect of $f_{soc}$ again shown in the background for each district. The thick black lines represent borders between federal states, while the thinner grey borders separate federal districts. Only the estimates of the quasi likelihood approach are shown.}
    	\label{fig:quasistart}
    \end{figure}

\begin{figure}[t!]
    	\centering
    	\includegraphics[width=0.75\linewidth]{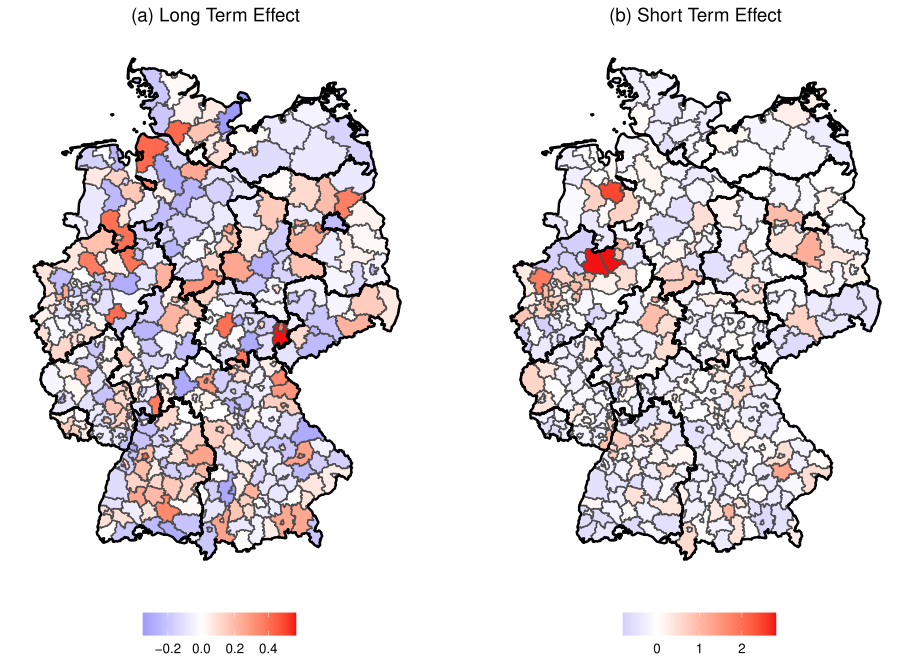}
    	\caption{(a): Maximum posterior modes of the long-term random effects $a_i$. (b) Maximum posterior modes of the short-term random effects $b_i$. The thick black lines represent borders between federal states, while the thinner grey borders separate federal districts. Only the estimates of the quasi likelihood approach are shown.}
    	\label{fig:quasiend}
    \end{figure}

    To exhibit the robustness of our findings, we compare the results under an alternative model specification based on the quasi-likelihood \citep{Wedderburn1974} combined with multiplicative random effects \citep{Firth1991}. In contrast to the main analysis, not the complete conditional distribution but solely its first two moments are specified. To enable a clear comparisons, we carry out the exact same procedure detailed in the main article but substitute the negative binomial likelihood conditional on $c$ for a conditional random variable $Y_{i,g,t} \mid x_{i,g,t-1}, y_{i,g,t-1}, a_i, b_i$ with $\mathbb{E}(Y_{i,g,t} \mid x_{i,g,t-1}, y_{i,g,t-1}, a_i, b_i) = \exp\lbrace \nu_{i,g,t}^{END} + \nu_{i,g,t}^{EPI} \rbrace$ and $\text{Var}(Y_{i,g,t} \mid x_{i,g,t-1}, y_{i,g,t-1}, a_i, b_i) = \exp\lbrace \nu_{i,g,t}^{END} + \nu_{i,g,t}^{EPI} \rbrace \phi$, where $\phi$ is an additional dispersion parameter. Consecutively, we again correct for the multiple imputation scheme and the estimates relating to the social activity during COVID-19 are given together with the original findings in Figure \ref{fig:quasi1} and \ref{fig:quasi2}. It is clearly visible that the findings hardly differ from the ones presented in the main article and that all conclusions drawn in the main article based on the negative binomial approach are consistent with the ones of the quasi likelihood model. The results of all other covariates are similar for the quasi-likelihood model to the results shown in the principal analysis (see Figure \ref{fig:quasistart} to \ref{fig:quasiend}).  
    
    \begin{figure}[t!]
    	\centering
    	\includegraphics[width=0.35\linewidth]{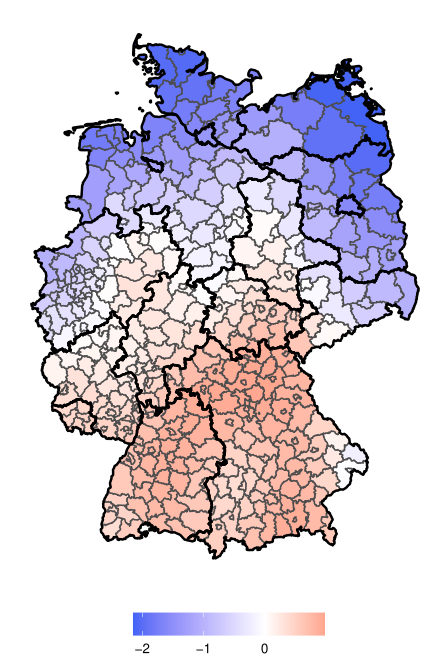}
    	\caption{Estimated smooth spatial effect $f_{coord}$. The thick black lines represent borders between federal states, while the thinner grey borders separate federal districts. Only the estimates of the quasi likelihood approach are shown.}
    \end{figure}

    \newpage
    \bibliographystyle{chicago}
    
    \bibliography{library_alt}